\documentclass[12pt]{article}
\usepackage[utf8]{inputenc}
\usepackage[usenames,dvipsnames]{color}
\usepackage{amsmath}
\usepackage{amssymb}
\usepackage{mathptmx}
\usepackage{authblk}

\usepackage[a4paper, margin=1in]{geometry}

\usepackage{multirow}
\usepackage{array}
\usepackage{rotating}
\usepackage{booktabs}%
\usepackage{romannum}

\usepackage[breaklinks,colorlinks,citecolor=RoyalBlue,linkcolor=magenta]{hyperref}

\usepackage[super,compress,comma,sectionbib]{natbib}
\usepackage{graphicx}
\usepackage{subcaption}
\usepackage[labelfont=bf]{caption}
\DeclareCaptionLabelSeparator{bar}{ \textbar{} }
\newcommand{\arcsec}{$^{\prime\prime}$}
\newcommand{\arcmin}{$^{\prime}$}
\newcommand{\arcdeg}{$^{\circ}$}

\newcommand\farcs{\mbox{$.\!^{\prime\prime}$}}

\def\jnl@style{\it}
\def\aaref@jnl#1{{\jnl@style#1}}

\newcommand{\araa}{Annu. Rev. Astron. Astrophys.}   
\newcommand{\aj}{Astron. J.}   
\newcommand{\apj}{Astrophys. J.}   
\newcommand{\apjl}{Astrophys. J. Lett.}   
\newcommand{\apjs}{Astrophys. J. Suppl. Ser.}   
\newcommand{\aap}{Astron. Astrophys.}   
\newcommand{\mnras}{Mon. Not. R. Astron. Soc.}   
\newcommand{\nat}{Nature} 
\newcommand{\physrep}{Phys. Rep.}   

\renewcommand{\figurename}{Fig.}

\newcommand{\mthc}{\mbox{CH$_3$CN}}

\newcommand{\tmthc}{\mbox{$^{13}$CH$_3$CN}}

\newcommand{\kms}{\mbox{km\,s$^{-1}$}}

\newcommand{\msun}{\mbox{$M_\odot$}}

\title{\textbf{Detection of a septuple stellar system in formation via disk fragmentation}}

\author[1,2,3*]{Shanghuo Li}
\affil[1]{\small School of Astronomy and Space Science, Nanjing University, Nanjing, China; \url{shli@nju.edu.cn}}
\affil[2]{Key Laboratory of Modern Astronomy and Astrophysics, Nanjing University, Ministry of Education, Nanjing, China}
\affil[3]{Max Planck Institute for Astronomy, Heidelberg, Germany}

\author[3]{Henrik Beuther}

\author[4,5]{Andr\'e Oliva}
\affil[4]{Space Research Center (CINESPA), School of Physics, University of Costa Rica, San Jos{\'e}, Costa Rica}
\affil[5]{Faculty of Physics, University of Duisburg-Essen, Duisburg, Germany}

\author[5]{Vardan G. Elbakyan}

\author[6]{Stella S. R. Offner}
\affil[6]{Department of Astronomy, The University of Texas at Austin, Austin, TX, USA}

\author[5]{Rolf Kuiper}

\author[1,2]{Keping Qiu}

\author[7]{Xing Lu}
\affil[7]{Shanghai Astronomical Observatory, Chinese Academy of Sciences, Shanghai, China}

\author[8]{Patricio Sanhueza}
\affil[8]{Department of Astronomy, School of Science, The University of Tokyo, Tokyo, Japan}

\author[9]{Huei-Ru Vivien Chen}
\affil[9]{Institute of Astronomy \& Department of Physics, National Tsing Hua University, Hsinchu, Taiwan}

\author[10]{Qizhou Zhang}
\affil[10]{Center for Astrophysics, Harvard \& Smithsonian, Cambridge, MA, USA}

\author[11,12,9]{Fernando A. Olguin}
\affil[11]{Center for Gravitational Physics, Yukawa Institute for Theoretical Physics, Kyoto University, Kyoto, Japan}
\affil[12]{National Astronomical Observatory of Japan, National Institutes of Natural Sciences, Tokyo, Japan}

\author[13,14]{Chang Won Lee}
\affil[13]{Korea Astronomy and Space Science Institute, Daejeon, Republic of Korea}
\affil[14]{University of Science and Technology, Daejeon, Republic of Korea}

\author[15]{Ralph E. Pudritz}
\affil[15]{Origins Institute and Department of Physics and Astronomy, McMaster University, Hamilton, Ontario, Canada}

\author[16]{Shuo Kong}
\affil[16]{Steward Observatory, University of Arizona, Tucson, AZ, USA}

\author[17]{Rajika L. Kuruwita}
\affil[17]{Heidelberg Institute for Theoretical Studies, Heidelberg, Germany}

\author[7]{Qiuyi Luo}

\author[12]{Junhao Liu}

\date{}
 
\begin{document}

\maketitle

{\color{blue}
\noindent
Stellar multiple systems play a pivotal role in cluster dynamics and stellar evolution, leading to intense astronomical phenomena like X-ray binaries, gamma-ray bursts, Type \Romannum{1}a supernova, and stellar mergers, which are prime sources of gravitational waves. 
However, their origin remains poorly understood.  
Here we report the discovery of a septuple protostellar system embedded in a Keplerian disk within the high-mass star-forming region NGC\,6334IN, with close separations of 181-461 AU. 
The stability analysis reveals that the disk surrounding the septuple system is dynamically unstable, indicating that the  septuple system formed via disk fragmentation. 
Previous studies have typically found only 2--3 members forming via disk fragmentation in both low- and high-mass star-forming regions. 
Our findings provide compelling observational evidence that the fragmentation of a gravitationally unstable disk is a viable mechanism for the formation of extreme high-order multiplicity, confirming what was previously only a theoretical concept. 
The results shed new light on the formation of extreme high-order multiplicity in cluster environments. 
}

Stellar multiple systems are believed to be a ubiquitous outcome of the star formation process~\citep{Zinnecker2007,Duchene2013,Reipurth2014,Offner2023}. 
Their complex dynamical interactions play a crucial role in shaping both cluster dynamics and stellar evolution~\citep{Portegies-Zwart2010,Sana2012,Marchant2024,Tauris2006,Tauris2017,Schneider2019,Mandel2022}. 
However, the dominant formation mechanism of multiplicity remains highly debated, with possibilities including disk~\citep{Bonnell1994,Bonnell1994b,Kratter2010,Oliva2020}, core~\citep{Larson1972,Offner2010,Li2024}, and filament fragmentation~\citep{Pineda2015}. 
Disk fragmentation has been proposed as a mechanism capable of forming  higher order multiplicity than core fragmentation \citep{Kratter2010, Oliva2020,Guszejnov2023}.
Observations of protostellar multiplicity suggest that triple and higher-order  systems (hereafter high-order system) are rare in low-mass protostars \citep{Pineda2015,Tobin2016,Tobin2022}.
High-order systems are believed to be more common in high-mass star formation regions \citep{Offner2023,Li2024}, as evidenced by the frequent detection of high-order main-sequence stellar systems in the Milky Way  \citep{Raghavan2010,Tokovinin2021}. 
However, observational studies on the formation of 
high-order multiple systems remain rare due to the direct imaging of such high-order systems being challenging for the early stages of high-mass star  formation \citep{Li2024,Beltran2016,Brogan2016,Beuther2017,Orozco-Aguilera2017,Ilee2018,Zhang2019,Guzman2020,Sanhueza2025,Suri2021,Cyganowski2022,Olguin2022,Kong2023}. 
Observing the initial configuration of multiple systems is key to unambiguously constraining  their origins and evolutionary paths, since dynamical interactions and evolution rapidly modify the initial separations~\citep{Lee2019,Kuruwita2023}.

\vspace{0.5em}
\noindent{\textbf{Discovery of the septuple system}}

\begin{figure}
\centering
\includegraphics[width=1.0\textwidth]{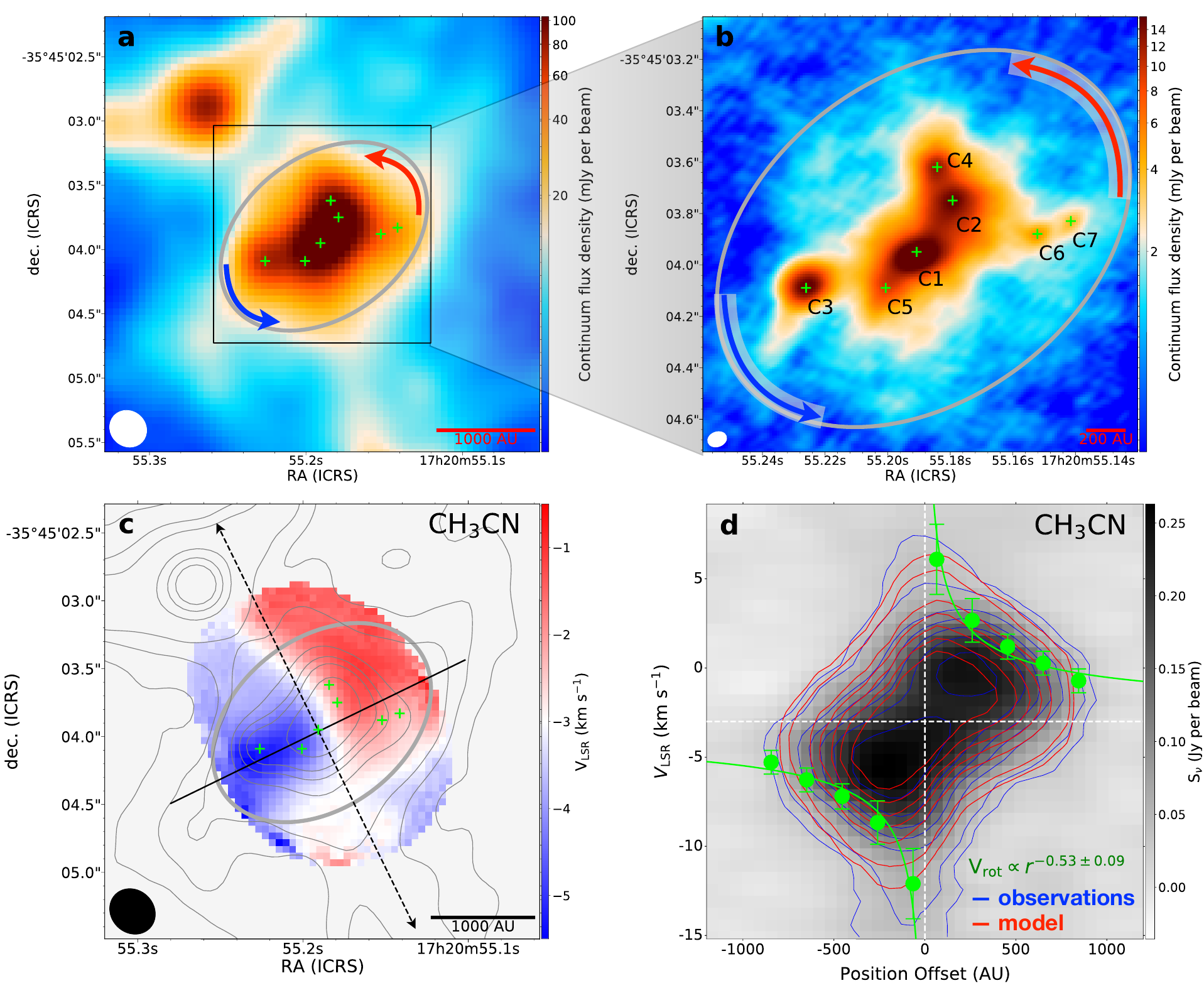}
\caption{
\textbf{Continuum images and gas kinematics.} 
\textbf{(a)} ALMA low-resolution ($\theta \sim$ 364~AU) 1.3~mm continuum image. 
\textbf{(b)} ALMA high-resolution ($\theta \sim$ 78~AU) 1.3~mm continuum image. Green pluses indicate the septuple system (C1 -- C7).
The gray ellipse represents the parent structure of the septuple system, while the blue and red arrows illustrate its rotational direction. 
\textbf{(c)} Intensity-weighted velocity map of CH$_{3}$CN $12_{4} -11_{4}$ is overlaid with 1.3~mm continuum as grey contours.  The contours are [5, 15, 25, 35, 80, 120, 160, 260]$\times \sigma$ with $\sigma$ of 0.36~mJy per beam for 1.3~mm continuum. 
The black dashed and solid lines present the directions of the SiO outflow and the cut line of position-velocity (PV) diagram.  
\textbf{(d)} PV diagram of CH$_{3}$CN $12_{4} -11_{4}$. 
Blue and red contours show the observations and model, respectively. 
Contour levels  are [5, 8, 17, 23, 29, 36, 44]$\times \sigma_{\rm rms}$ interval, where $\sigma_{\rm rms}$ is 0.05 Jy per beam.  
The green circles show the best-fit rotation velocities at each radius derived from {\tt 3DBarolo} (\hyperref[sec:methods]{Methods}).  
The green solid curve indicates the best-fit power-law, 
$V_{\rm rot} \propto r^{-0.53 \pm 0.09}$, to rotation velocities.  
The horizontal dashed line shows the systemic velocity of  3~\kms. 
The ellipses in the bottom-left corner of each panel denote the synthesized beam of the corresponding image.  
The linear scale bar is displayed in the bottom-right corner of the panels.
}
\label{fig:cont0}
\end{figure}

Using the Atacama large millimetre/submillimetre array (ALMA), we observed the massive protocluster NGC\,6334I-North (hereafter NGC\,6334IN) with a resolution of $\sim$0.06\arcsec\ (equivalent to 78~AU at the source distance of 1.3 kpc)~\cite{Chibueze2014,Reid2014}  and 0.28\arcsec\ (equivalent to 364~AU) at the wavelength of 1.3~mm (\hyperref[sec:methods]{Methods}). 
In the high-resolution 1.3~mm continuum image, we detect a gravitationally bound multiple system  composed of  seven condensations (Fig.~\ref{fig:cont0} and Extended Data Fig.~\ref{fig:cont1}; \hyperref[sec:methods]{Methods}), hereafter referred to as the septuple system. 
This represents an extremely high-order multiple system identified so far in high-mass star-forming regions.

The detected condensations have projected radii between 89 and 189 AU (Table.~\ref{tab1}). The projected separations of the nearest-neighbor pairs range from 181 to 461 AU, with an average separation of 298 AU. The measured separations are consistent with the characteristic separation ($\lesssim$~500 AU) of multiplicity formation via fragmentation of a large gravitationally unstable accretion disk  \cite{Offner2023,Oliva2020,Machida2008,Meyer2019}, and are smaller than those typically formed via core fragmentation ($\gtrsim$~1000~AU)~\cite{Li2024,Lee2019,Kuruwita2023,Offner2016}. 
The estimated condensation gas masses ($M_{\rm gas}$) of the septuple system members are between 0.04~\msun{} and  0.25~\msun{} based on the 1.3~mm thermal dust emission (\hyperref[sec:methods]{Methods}). 
The derived masses are considered as lower limits due to the optical depth effects and missing flux in the interferometric data. 
The condensations likely harbour protostellar object(s), as suggested by their gravitationally bound nature (\hyperref[sec:methods]{Methods}).

\vspace{0.5em}
\noindent{\textbf{Properties and fragmentation of the circum-septuple disk}}

The septuple system is embedded in a single parent structure, as revealed by the ALMA low-resolution data (Fig.~\ref{fig:cont0}a). The parent structure shows an obvious velocity gradient in the northwest-southeast direction (Fig.~\ref{fig:cont0}c), which is perpendicular to the outflow direction running from northeast to southwest (Extended Data Fig.~\ref{fig:kine}). 
In addition, the position-velocity (PV) diagram of \mthc, which is a typical disk tracer \cite{Johnston2015}, is consistent with the velocity profile of a Keplerian rotation  (Fig.~\ref{fig:cont0}d). 
This evidence indicates that the septuple system is embedded in a Keplerian-like disk structure.

We constrain the three-dimensional (3D) configuration of the disk based on the gas kinematics (\hyperref[sec:methods]{Methods}). 
The best-fitted inclination angle ($i$) of the disk is 26.3$^{\circ}$ (0$^{\circ}$ for face-on). 
The rotation velocities in the disk decrease as a function of disk radius and are well described by a power-law profile, $V_{\rm rot} \propto r^{-0.53 \pm 0.09}$, within a radius of 850 AU (Fig.~\ref{fig:cont0}d). 
The derived velocity profile is consistent with the Keplerian rotation of  $V_{\rm rot} \propto r^{-0.5}$.
The disk has a gas mass $M_{\rm disk}$ of 1.87~\msun\  and a projected radius of 885~AU based on dust continuum emission (\hyperref[sec:methods]{Methods}).
The enclosed mass $M_{\rm cl}$ calculated from fitting a Keplerian model to the disk rotation velocities is 7.6 $\pm$ 1.4~\msun\ (\hyperref[sec:methods]{Methods}). 
The protostellar mass, estimated as $M_{\ast}$ = $M_{\rm cl}-M_{\rm disk}-\sum M_{\rm gas}$ = 4.89~\msun, results in a disk-to-star mass ratio of $M_{\rm disk}/M_{\ast} = 0.2$--$0.6$, accounting for uncertainties in the disk mass, condensation gas mass, and protostellar mass.
This ratio exceeds the approximate minimum ratio of 0.1 required to  trigger gravitationally instability~\cite{Lodato2004,Kratter2016}.  
The derived $M_{\rm disk}/M_{\ast}$ ratio is similar to those of gravitationally unstable disks observed in other high-mass star-forming regions, e.g., G353.273 (ref.~\cite{Motogi2019}).  
The high $M_{\rm disk}/M_{\ast}$ ratio suggests that the disk is in a gravitationally unstable regime and has the potential to develop additional fragments as it evolves.

\begin{figure}[!h]
\centering
\includegraphics[width=1\textwidth]{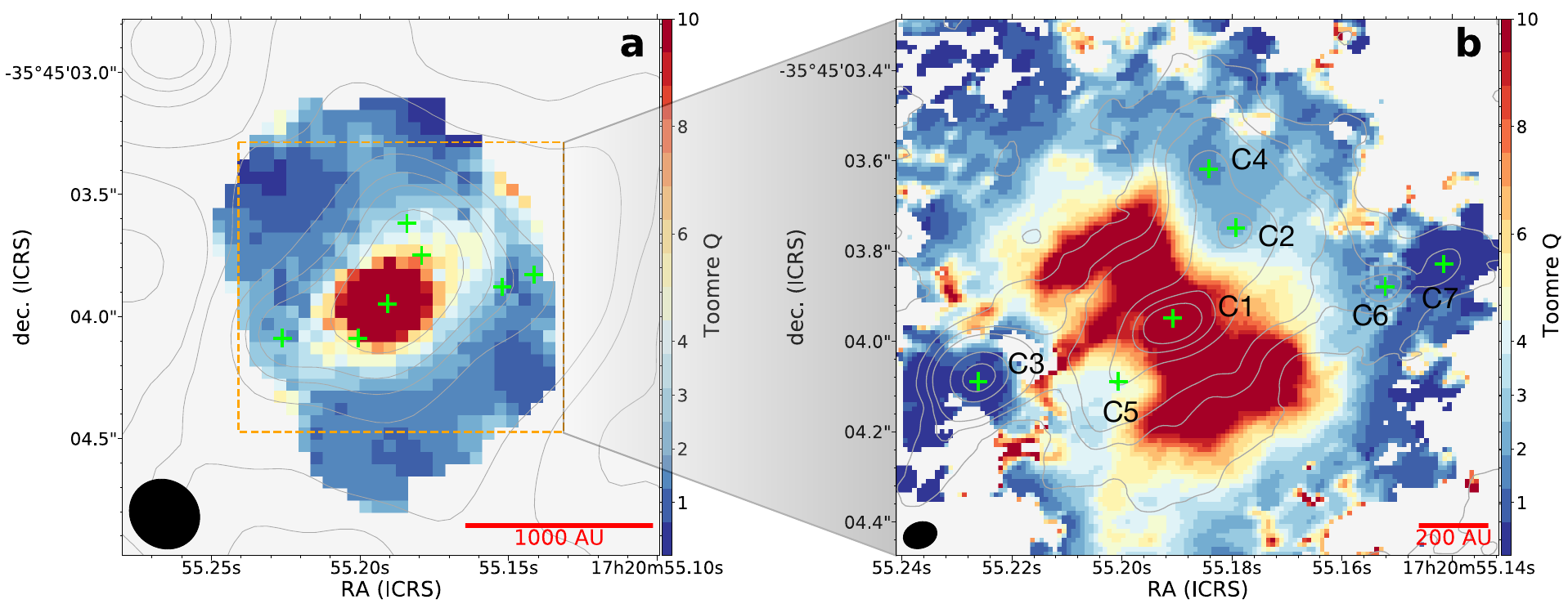}
\caption{
\textbf{Toomre $Q$ parameter in the disk.} 
\textbf{(a)} and \textbf{(b)} The Toomre $Q$ parameter maps determined with the low-resolution and high-resolution data, respectively. The Toomre $Q$ parameters reveal that the disk around condensations C2, C3, C4, C5, C6, and C7 is marginally unstable. 
The grey contours show the continuum image, with contour levels are [5, 15, 25, 35, 80, 120, 160, 260]$\times \sigma$ with $\sigma$ of 0.36~mJy per beam for the low-resolution image, and [15, 25, 30, 35, 70, 85, 100, 120, 160, 200] $\times \sigma$ with $\sigma$ of 0.12~mJy per beam for the high-resolution image. 
The green pluses indicate the condensations. 
The black ellipses in the bottom-left corner of each panel denote the synthesized beam of corresponding images.  The linear scale bar is presented in the bottom-right corner of the panels.  
}
\label{fig:TnQ}
\end{figure}

The Toomre $Q$ parameter is widely used to quantify the gravitational instability of a self-gravitating disks \cite{Toomre1964}, where $Q \lesssim$ 1--2 indicates instability \cite{Kratter2016,Durisen2007}. 
In Fig.~\ref{fig:TnQ}, we show Toomre $Q$ parameter maps derived from the low-resolution and high-resolution data (\hyperref[sec:methods]{Methods}).  
The figure shows that the Toomre $Q$ value generally decreases with increasing disk radius in both low-resolution and high-resolution maps (Fig.~\ref{fig:TnQ} and Extended Data Fig.~\ref{fig:radTnQ}). 
The central region shows a high value of $Q >$ 10, suggesting that the region is gravitationally stable. 
This stability could be attributed to the protostellar radiation, which raises the local temperature and thus may stabilize the central region of the disk. 
On the other hand, the outer disk has relatively low $Q$ values ($\lesssim$ 3), indicating that the outer disk is marginally unstable ($Q \approx$ 1--2) given that the uncertainty is about 54\% in $Q$ (\hyperref[sec:methods]{Methods}). 
As expected, low Q values (mean value $\leqslant$2.7) are revealed at the locations of  condensations C2, C3, C4, C6, and C7, which have likely already collapsed to form protostars (Fig.~\ref{fig:TnQ}b). 
Condensation C5 is located at a position with a higher $Q$ value of 4.8, while it remains around the critical value  when accounting for uncertainty. 
The marginal instability at these positions suggests that disk fragmentation is responsible for the formation of this high-order multiple system.  
We note that the estimated $Q$ values are likely upper limits because  unresolved substructures, such as spiral arms, can increase the column densities, and consequently result in locally reduced $Q$ values. 
The derived Toomre $Q$ parameter is consistent with the high $M_{\rm disk}/M_{\ast}$ ratio of 0.2--0.6, placing the disk in a gravitationally unstable regime. 
These results further support the scenario that the disk is dynamically unstable, leading to disk fragmentation that produces the septuple system.

The measured separations can be deprojected by applying a correction factor of 1/cos$(i)$, where $i$~=~26.3$^{\circ}$ is the derived disk inclination angle. 
The deprojected separations of the septuple system, with a mean separation of 332~AU, are notably smaller than the typical separation of $\gtrsim$1000~AU for the multiple system formed through core fragmentation~\cite{Li2024,Lee2019,Kuruwita2023,Offner2016}. 
In addition, the separations of the septuple system are 3.5 times smaller than the Jeans length of 1168~AU for thermal Jeans fragmentation, assuming their parent structure is an approximately spherical core structure (\hyperref[sec:methods]{Methods}).  
On the other hand, the expected Jeans length is up to 3460 AU if one includes the isotropic turbulence support (\hyperref[sec:methods]{Methods}). 
Therefore, the measured separations of the septuple system are inconsistent with Jeans fragmentation of the dense core.  
In addition, core fragmentation tends to produces multiple systems with a wide separation of  $\gtrsim$1000~AU,  although inward migration may occur over time~\cite{Guszejnov2023,Lee2019,Kuruwita2023}. 
Furthermore, the observed kinematics are consistent with formation via disk  fragmentation. 
These results suggest that core and filament fragmentation are unlikely to be responsible for the formation of the observed septuple system, given its early evolutionary stage and relatively isolated from nearby sources (Extended Data Fig.~\ref{fig:cont1}). 
The septuple system is gravitationally bound, with the outermost condensation of C7 being the most strongly gravitationally bound member (Extended Data Fig.~\ref{fig:EiWi}). 
Overall, the results demonstrate that the septuple system is the outcome of disk fragmentation. 
Given the early age of the system and its location in a dense protocluster forming region, it is likely that the multiplicity of the resulting star cluster will be affected by mergers or (proto)stellar captures~\cite{Oliva2020,Guszejnov2023,Lee2019,Kuruwita2023}.

\begin{figure}[!h]
\centering
\includegraphics[width=1\textwidth]{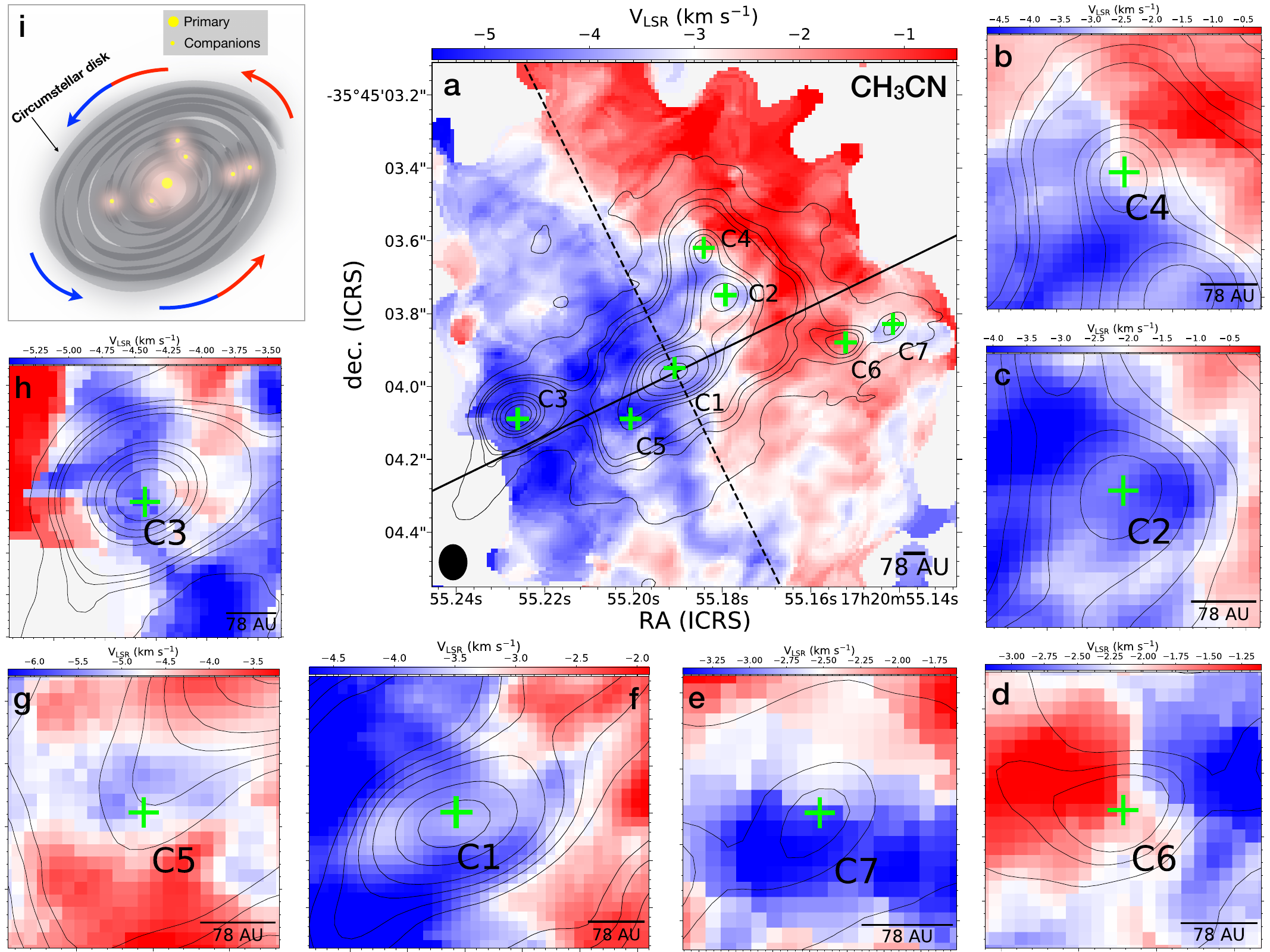}
\caption{
\textbf{Intensity-weighted velocity map of \mthc{} within the disk.} 
\textbf{(a)} Intensity-weighted velocity map of \mthc{} $12_{4} -11_{4}$ derived from the high-resolution data. 
The black contours show the high-resolution 1.3~mm continuum image, with contour levels at [15, 25, 30, 35, 70, 85, 100, 120, 160, 200] $\times \sigma$, where $\sigma$ is 0.12~mJy per beam. 
The synthesized beam of \mthc{} image is shown in the bottom-left corner as a black ellipse. The black dashed and solid lines indicate the directions of the SiO outflow and the cut line of PV diagram.
\textbf{(b)--(h)} Zoom-in views of the velocity map for each condensation. 
The linear scale bar is presented in the bottom-right corner of each panel.  
\textbf{(i)} A schematic illustration of the septuple system in a rotating circumstellar disk. 
}
\label{fig:vlsr}
\end{figure}

\begin{figure}
\centering
\includegraphics[width=1\textwidth]{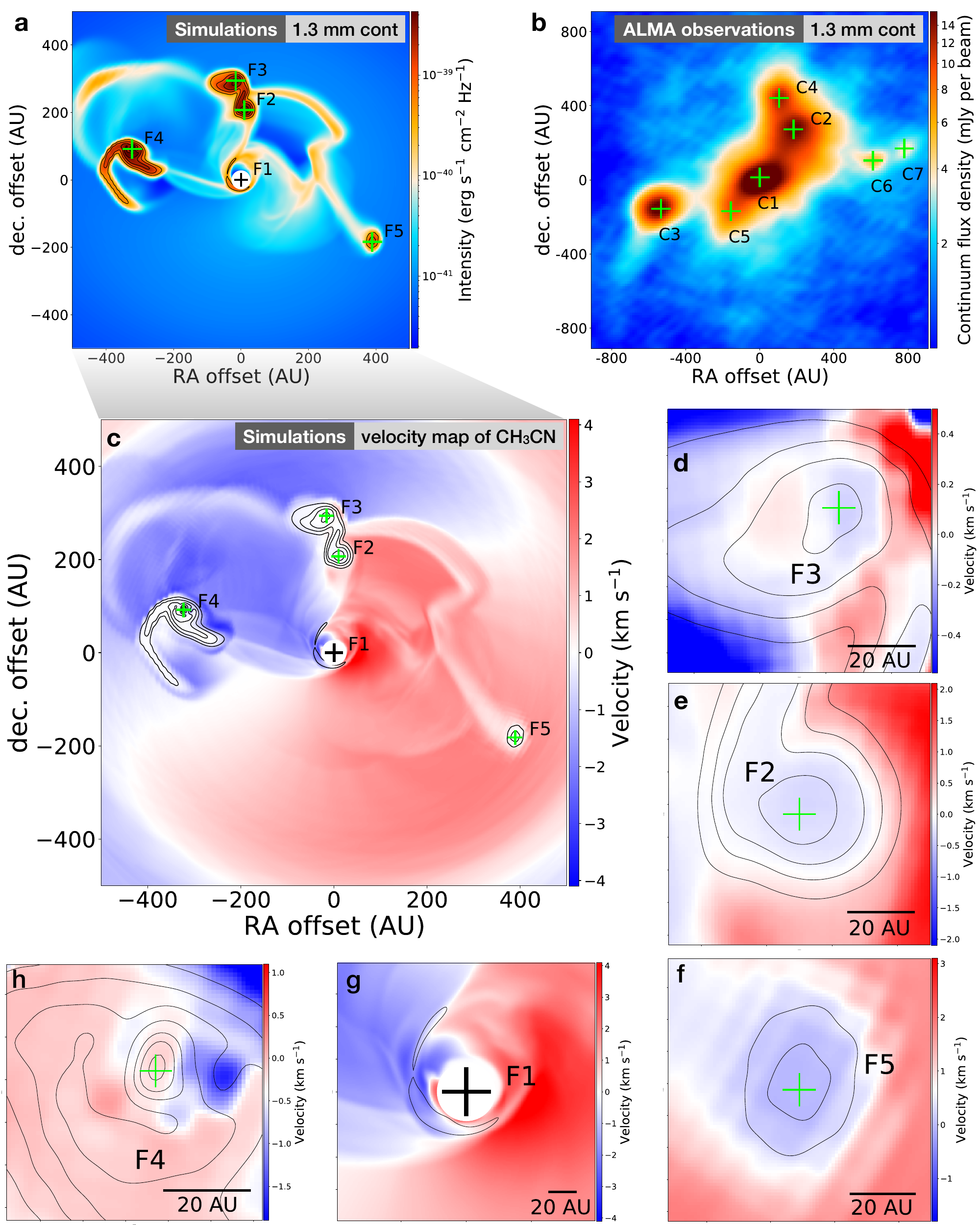}
\caption{\textbf{1.3~mm continuum and velocity maps of simulations.} 
\textbf{(a)} The 1.3~mm continuum image of the model simulations is inclined by 27$^{\circ}$ at a distance of 1.3 kpc.  The central source is masked out. 
The black contour levels are [0.45, 1.18, 2.25, 3.62, 5.35, 6.59, 7.83, 9.08]$\times 10^{-39}$ erg s$^{-1}$ cm$^{-2}$ Hz$^{-1}$. 
\textbf{(b)} The ALMA 1.3mm continuum image. 
The simulations image reveals a configuration of a multiple system similar to the observations, though the disk size in the simulations appears slightly smaller compared to NGC\,6334IN.  
\textbf{(c)} Intensity-weighted velocity map of CH$_{3}$CN $12_{4} -11_{4}$ for the mode simulations at an inclination of 27$^{\circ}$. 
\textbf{(d)}--\textbf{(h)} The zoomed views of the velocity distributions around each fragment (F1--F5).  
The velocity profiles in both the large-scale disk and the fragments resemble those seen in NGC\,6334IN. 
}
\label{fig:simul}
\end{figure}

\vspace{0.5em}
\noindent{\textbf{Kinematics of the septuple system}}

We further investigate the gas kinematics of each condensation within the circum-septuple disk using the velocity map of  \mthc{}  obtained from the high-resolution data (Fig.~\ref{fig:vlsr}). 
The velocity map shows prominent variations throughout the disk, with a general large-scale velocity gradient oriented along the northwest-southeast direction, consistent with observations from the low-resolution data (Fig.~\ref{fig:cont0}c). 
A clear velocity gradient is observed across some condensations, resembling those seen in simulations of multiple systems formed via disk fragmentation~\cite{Oliva2020,Ahmadi2019} (Fig.~\ref{fig:simul}). 
The velocity gradients around the condensations are generally consistent with those of their parent disk along the northwest-southeast direction, with the exception of C5 and C6, where the velocity gradients are oriented along the west-east and north-south directions, respectively.  
This indicates that C5 and C6 may have decoupled from the larger flow and experienced dynamical interactions that changed their kinematics. 
The \mthc{} emission shows enhanced integrated intensity and broadened intensity-weighted linewidth around the condensations (Extended Data Fig.~\ref{fig:mom02}), with the one exception of C7, which has no significant variation compared to the surrounding region.  
This could be due to C7 having a relatively lower temperature and mass than the other companions (Extended Data Fig.~\ref{fig:Tgas}), resulting in undetectable variations given the current sensitivity.

Motivated by our observations, we present a simulation that models the formation of multiplicity via disk fragmentation in a high-mass star-forming region (\hyperref[sec:methods]{Methods}). 
As shown in Fig.~\ref{fig:simul}a, five bright fragments emerge within the simulated disk, showing a multiple system configuration similar to that observed in NGC\,6334IN (Fig.~\ref{fig:simul}b), although the simulated disk is slightly smaller in size. 
Additional fragments will form as the system evolves \cite{Oliva2020}, increasing the complexity of the multiplicity (Extended Data Fig.~\ref{fig:simu7}). 
The central protostar has a mass of $\sim$6.7~\msun{}, closely matching  the measured protostellar mass of $M_{\rm \ast}$ = 5.7~\msun. 
Assuming a dust temperature of 200 K and following the same procedures as for the observations, 
the fragment gas masses are estimated to range from 0.03 to 0.23 \msun{}, which is comparable to the condensation gas masses of the septuple system members in NGC\,6334IN. 

Therefore, disk fragmentation around a forming central star can lead to formation of high-order multiple systems, such as the one we discovered in NGC\,6334IN. 
The velocity maps of \mthc{} from the simulations show obvious velocity gradient across the disk and some embedded fragments (Fig.~\ref{fig:simul}c--h),  resembling the gas kinematics observed in NGC\,6334IN (Fig.~\ref{fig:vlsr}). 
The consistency between the simulations and observations in the continuum emission and kinematic features indicates that disk fragmentation is responsible for the formation of the septuple system in NGC\,6334IN.

\vspace{0.5em}
\noindent{\textbf{Implications for extreme high-order multiplicity formation}}

The discovery of a septuple protostellar system embedded in a parent Keplerian-like disk toward NGC\,6334IN provides the direct observational evidence that disk fragmentation, driven by gravitationally instability, can lead  to the formation of a septuple system within a 1500~AU scale. 
This finding highlights the crucial role of disk fragmentation in the formation of extreme high-order multiplicity during cluster formation. 
The results confirm that disk fragmentation generate higher-order multiplicity on smaller scales than core fragmentation, providing a unique benchmark for existing multiplicity formation theories, which are critical for understanding the formation and evolution of high-mass stellar clusters.

%
\clearpage
\begin{sidewaystable}
\caption{Properties of condensations}\label{tab1}
\setlength\tabcolsep{2pt} 
\resizebox{1\textwidth}{!}{%
\scriptsize 
\begin{tabular*}{\textheight}{@{\extracolsep\fill}lcccccccccccccccccc}
\toprule%
Condensation   & RA    &  DEC & $I_{\rm peak}$ & $S_{\nu}$ &  size  & Radius & 
$T_{\rm gas}$  & $Q$  & $M_{\rm gas}$   & $N_{\rm H_2}$&  $\sum$ &  $\rho$  &  $V_{\rm LSR}$ &  $W_{\rm i}$&  $E_{i}$ &  $\alpha_{\rm vir}$ \\
  & (hh:mm:ss.sss) & (dd:mm:ss.sss) & (mJy per beam) & (mJy) &  (\arcsec$\times$\arcsec, $^{\circ}$)   & (\arcsec [au]) & 
(K)   & & ($M_\odot$)   & (cm$^{-2}$)  &  (g cm$^{-2}$)   &  (cm$^{-3}$ ) 
&  (km s$^{-1}$)  &  ($M_{\odot}$ km$^{2}$ s$^{-2}$) &  ($M_{\odot}$ km$^{2}$ s$^{-2}$)  & \\   
  &  &  &  &  &  &  & 
  & & & ($\times 10^{24}$)  & &  ($\times 10^{9}$) 
& & &  & \\    \hline
(1) & (2) & (3) &  (4)  &  (5)  &  (6)    & (7)    & (8) & (9) & (10) & (11) & (12) & (13) &  (14)  & (15) & (16)  & (17)\\  \hline
C1 & 17:20:55.191 &  -35:45:03.949  & 28.03$\pm$1.02 & 162.80$\pm$5.93 & 0.192 $\times$ 0.110, 107 & 0.145[189] & 377 & 28.5 & 0.251  & 6.83 & 19.84 & 1.12 & -3.28$\pm$0.31 &  ... &  ... & 1.0 \\
C2 & 17:20:55.179 &  -35:45:03.749  & 21.22$\pm$1.12 & 93.45$\pm$4.94 & 0.129 $\times$ 0.116, 109 & 0.123[159] & 416 & 2.7 & 0.131  & 4.38 & 14.53 & 0.98 & -0.91$\pm$0.45 & -1.96  & 0.86 & 1.2 \\ 
C3 & 17:20:55.226 &  -35:45:04.089 & 19.64$\pm$1.88 & 65.20$\pm$6.24 & 0.119 $\times$ 0.086, 126 & 0.101[131] & 188 & 0.4 & 0.204  & 11.47 & 33.44 & 2.72 & -4.45$\pm$0.19 & -1.66  & 0.65  & 0.2 \\
C4 & 17:20:55.184 &  -35:45:03.619 & 14.45$\pm$1.18 & 47.73$\pm$3.89 & 0.122 $\times$ 0.081, 151 & 0.099[129] & 401 & 2.0 & 0.069  & 3.50 & 11.71 & 0.97 & -1.76$\pm$0.29 & -0.75  & 0.16  & 0.5 \\ 
C5 & 17:20:55.201 &  -35:45:04.089 & 11.82$\pm$1.16 & 37.92$\pm$3.71 & 0.130 $\times$ 0.072, 145 & 0.096[125] & 301 & 4.8 & 0.074  & 4.02 & 13.26 & 1.13 & -5.49$\pm$0.33 & -1.46  & 0.68 & 1.8  \\
C6 & 17:20:55.152 &  -35:45:03.879 & 6.42$\pm$1.33 & 17.91$\pm$3.71 & 0.112 $\times$ 0.066, 58 & 0.086[111] & 260 & 1.9 & 0.040  & 2.21 & 9.19 & 0.88 & -1.58$\pm$0.44 & -0.31  & 0.12  & 0.6 \\ 
C7 & 17:20:55.142 &  -35:45:03.829 & 4.86$\pm$1.35 & 9.99$\pm$2.78 & 0.078 $\times$ 0.060, 117 & 0.068[89] & 87 & 0.3 & 0.070  & 5.29 & 25.02 & 3.01 & -3.04$\pm$0.51 & -0.42  & 0.0002  & 0.1 \\  \hline
\end{tabular*}}
\footnotetext{Note: Columns (1)--(3) present the name, right ascension, and declination of condensations. 
The continuum peak intensity, flux density, beam-deconvolved size, and beam-deconvolved radius are shown in columns (4)--(7).  
Column (8) is the condensation averaged gas temperature derived from \mthc{}.  Column (9) is the averaged Toomre $Q$ parameter. 
The condensation gas mass, the averaged H$_2$ column density, averaged surface density, averaged volume density, and centroid velocity determined by \tmthc{} $13_{3}-12_{3}$ are shown in columns (10) -- (14).  
Columns (15) and (16) present the gravitational potential of the parent disk and kinetic energies of the condensations, respectively. 
The size and radius of condensations are observed values, which are not corrected by the inclination angle. 
The radius is equivalent to $\sqrt{\rm FWHM_{\rm maj} \times FWHM_{\rm min}}$. 
Column (17) is the estimated virial parameter, calculated under the assumption that the total condensation masses are 10 $M_{\rm gas}$.
}
\end{sidewaystable}

\clearpage

\phantomsection
\label{sec:methods}
\noindent{\large \textbf{Methods}}

\noindent\textbf{High-mass protocluster NGC\,6334IN}

NGC\,6334IN is a typical high-mass  protostellar cluster forming region associated with weak emission at 70~$\mu$m~\cite{Hunter2014,Konig2017,Sadaghiani2020,Cortes2021}, located at the north end of the NGC\,6334 molecular cloud complex that is a nearby (1.3 kpc) high-mass  star formation region encompassing sites at various evolutionary stages~\cite{Chibueze2014,Reid2014,Straw1989}. 
The NGC\,6334IN cloud contains a mass reservoir of $\sim$1000~\msun{} and has a luminosity of $1 \times 10^{4}\, L_{\odot}$ within a radius of 0.3~pc (ref.~\cite{Konig2017}). 
Previous ALMA and Submillimeter Array (SMA) observations of NGC\,6334IN have achieved the highest angular resolution of $\sim$0.5--0.6\arcsec{}(ref.~\cite{Hunter2014,Liu2023}). 
The SMA1 region, where the disk is located, contains a mass of $\sim$13 \msun{} based on the 1.3~mm continuum image obtained from SMA observations \cite{Brogan2009}. This should be considered as a lower limited due to the interferometer observations suffering from missing flux.
The presence of Class~I CH$_{3}$OH and H$_{2}$O masers suggests intense star formation activity toward the circumstellar disk in NGC\,6334IN~\cite{Brogan2009,Walsh1998}. 
As a representative relatively nearby example of a high-mass star-forming region, NGC\,6334IN serves an ideal laboratory for studying the formation of multiple systems in cluster environments.

\noindent\textbf{Observations and data reduction}

Observations were conducted as part of the Revealing the Origin of Multiplicity in high-mass  protostellar cluster with ALMA (ROMA) survey (Project ID: 2021.1.00713.S and 2022.1.\\00671.S; PI: Shanghuo Li). 
We used similar ALMA configurations to carry out the observations and adopted a similar approach to analyse the data as described in our previous work ref. \cite{Li2024}. 
The observations of NGC\,6334IN were performed with ALMA in Band 6  (at the wavelength of 1.3~mm) using the 12-m array. The short-baseline (or low-resolution) observations,  conducted on August-09-2022, utilized 44 antennas in a configuration similar to C40-5, while the long-baseline observations, carried on Jun-20-2023 and July-03-2023, employed 42 antennas in a configuration similar to C43-8. 
The baseline lengths are 15--1301~m and 84--8547~m for short-baseline and long-baseline observations, respectively. The correlator setup included four spectral windows centered at  217.806, 220.577, 232.194, and 234.314 GHz with a spectral resolution of 976.6 KHz ($\sim$1.3~\kms) and a bandwidth of 1.875 GHz.    
The total on-source time for NGC\,6334IN is 2~minutes for short-baseline observations and 15~minutes for long-baseline observations.   
The reference phase centre is ($\alpha$(ICRS), $\delta$(ICRS)) = 
(17h20m54.948s, $-$35\arcdeg{}45\arcmin{}07\farcs{429}).

The data were calibrated and imaged using the  CASA  (version 6.4.1-12) software package~\cite{McMullin2007}. 
Bandpass and flux calibrations were conducted with J1617--5848, and phase calibration was performed with J1713--3418. 
Line-free channels were used to generate the continuum data and produce continuum-subtracted data cubes for each observational epoch. 
Phase only self-calibration was performed  using the continuum data for short-baseline and long-baseline observations separately. 
The shortest time steps for phase self-calibration are 6 and 12 seconds for the short-baseline and long-baseline dataset, respectively.   
The self-calibration solutions were applied to the corresponding data cubes. 
We combined the short-baseline and long-baseline self-calibrated data (hereafter combined or high-resolution data) to recover extended emission for both continuum and data cubes. 
The short-baseline and combined data were imaged separately. 
We used the TCLEAN task with Briggs weighting and a robust parameter of 0.5 to image the continuum. The resultant continuum images have a synthesized beam of 0.30\arcsec $\times$ 0.27\arcsec{} (390~AU $\times$ 351~AU, Fig.~\ref{fig:cont0}a) with a position angle (P.A.) of 46.9\arcdeg{} and 0.076\arcsec $\times$ 0.058\arcsec{} (99~AU $\times$ 75~AU, Fig.~\ref{fig:cont0}b) with a P.A. = -71.1\arcdeg{} for short-baseline and combined dataset, respectively. The achieved 1$\sigma$ rms noise levels continuum images are about 0.36 mJy per beam and 0.12 mJy per beam for short-baseline and combined data, respectively. 
We also imaged the continuum using different weighting schemes with robust parameters of 0.5 and -2 for combined and long-baseline data only, resulting in slightly different beam size images (Extended Data Fig.~\ref{fig:cont1}). 
Data cubes for each spectral window were produced using the automatic masking procedure YCLEAN~\cite{Contreras2018}, which automatically cleans each map channel with customized masks.  A robust parameter of 0.5 is used to image the data cubes.  
More details on the YCLEAN algorithm can be found in ref.~\cite{Contreras2018}. 
The line images have 1$\sigma$ rms noise about 10~mJy per beam  and 3~mJy per beam with a channel width of $\sim$0.65~\kms{} for the short-baseline and combined data, respectively. 
To improve the signal-to-noise (SNR) ratio for the combined data, we used a robust parameter of 2 to image the \mthc{} lines, resulting in a  1$\sigma$ rms of 5~mJy per beam, where the beam size is 0.127\arcsec $\times$ 0.102\arcsec{} with a P.A. = -59.1\arcdeg{}.
The largest recoverable angular scales are 3.0\arcsec{} for short-baseline and combined data, as determined by the short-baselines in the array.

\vspace{0.2em}
\noindent\textbf{Condensation identification}

Here, the term condensations refers to substructures within a disk. 
We adopted a similar approach to identify the dense structure from the continuum images as described in our previous work ref.~\cite{Li2024}.  
To ensure reliable detection and avoid suspicious sources caused by diffuse emission in the combined data, we employed the CASA-{\tt imfit} task to preselect condensations from the continuum image generated using a uv range of $>$1.3 km (Extended Data Fig.~\ref{fig:cont1}a). We fitted four components to extract C1, C2, C4, and C5, one component was used for C3, and two components were used for C6 and C7. 
We then use the parameters of the preselect structures as input to CASA-{\tt imfit} for more accurate measurement of their parameters from the full combined continuum image (Fig.~\ref{fig:cont0}b),  including peak flux ($I_{\rm peak}$), integrated flux ($S_{\nu}$),  major and minor axis sizes (full width at half maximum; FWHM$_{\rm maj}$ and FWHM$_{\rm min}$), and P.A (Table.~\ref{tab1}). 
We reveal 7 condensations in the continuum image. 
Fig.~\ref{fig:cont0} and  Extended Data  Fig.~\ref{fig:cont1} show the identified condensations overlaid on 1.3~mm continuum images at different angular resolutions. 
Additionally, we also used the CASA-{\tt imfit} task to extract the parent structure of these condensations from the low-resolution 1.3~mm continuum image (Fig.~\ref{fig:cont0}).

\vspace{0.2em}
\noindent\textbf{Estimate of gas temperatures}

We derived the gas temperatures ($T_{\rm gas}$) map by modeling the $K$ = 0, 1, 2, 3, 4, 5, 6 ladders of the \tmthc{} $J=13\text{--}12$ transition in each spatial pixel of the ALMA low-resolution and high-resolution image cubes using the {\tt XCLASS} package~\cite{Moller2017}. 
An example spectrum of  \tmthc{} is presented in Supplementary Fig.~\ref{fig:13CH3CN}. 
The isotopologue \tmthc{}  lines have lower optical depth than those of \mthc{}, and thus the \tmthc{} better traces the interior of the structure, while \mthc{} is preferentially tracing the surface. 
To improve the fitting quality for the \tmthc{} lines, we included  CH$_{3}$COOH, CH$_{3}$OCHO, and CH$_{3}$OCH$_{3}$ lines during the fitting.     
For the low-resolution data, the derived rotational temperatures range from 40 to 600 K (Extended Data Fig.~\ref{fig:Tgas}a), which is used to estimate the mass and sound speed of the disk. 
For the high-resolution data, the derived rotational temperatures vary from 50 to 786~K  (Extended Data Fig.~\ref{fig:Tgas}b), which is used to calculate the condensation gas mass and the sound speed within the disk. 
The temperatures derived from the high-resolution data are in general similar to those from low-resolution data,  whereas the former shows greater small-scale variations across the disk. 
The derived high-resolution temperatures appear to decrease with distance from the centers of condensations, indicating internal heating, except for C7 that shows no significant variation. 
While optical depth and chemical processes might influence the estimated temperatures, this is the best estimate of the  temperature structure of the disk we can obtain.

\vspace{0.2em}
\noindent\textbf{Estimating gas mass from dust continuum emission}

The observed 1.3~mm continuum emission should be dominated by  thermal dust emission as the hydrogen recombination line (i.e., H30$\alpha$) is not detected. 
We calculate the gas mass, volume density ($\rho$), and surface density ($\sum$) for the condensations and their parent disk mass  following 
\begin{equation}
\label{Mgas}
M_{\rm gas} = \eta \frac{S_{\nu} \, \rm{D^2}}{\kappa_{\nu} \, B_{\nu}(T_D)},
\end{equation}

\begin{equation}
\label{density}
\rho =  \frac{M_{\rm gas}}{\frac{4}{3}\, \pi \, R^{3} \,\mu\, m_{\rm H}},
\end{equation}

\begin{equation}
\label{surface}
\sum = \eta \frac{I_{\nu}}{\kappa_{\nu}\, \Omega_{\rm beam} \, B_{\nu}(T_D)},
\end{equation}
where $\eta$ = 100 is the gas-to-dust ratio, 
$S_{\nu}$ is the measured integrated source flux,  
$I_{\nu}$ is the measured peak flux density,  
 $m_{\rm H}$ is the mass of a hydrogen atom, 
 $\mu$ = 2.8 is the mean molecular weight of the interstellar medium,  
D = 1.3~kpc is the distance to the source, 
$B_{\nu}$ is the Planck function at the dust temperature $T_D$,  
$\kappa_{\nu}$ is the dust opacity at frequency $\nu$, 
$\Omega_{\rm beam}$ is the solid angle of the beam, 
and $R$=$\sqrt{\rm FWHM_{\rm maj} \times FWHM_{\rm min}}$ 
is the radius of the condensation. 
Since NGC\,6334IN is still in a very early evolutionary stage 
of star formation,  we adopted a value of 0.9 cm$^{2}$ g$^{-1}$ for $\kappa_{\rm 1.3 mm}$, which is an appropriate assumption for a disk in the early phase of the star formation process~\cite{Tobin2016}.  
This value is retrieved from column 6 of Table 1 in ref.~\cite{Ossenkopf1994}. 
The typical disk density is sufficiently high ($\gg 10^{5}$ cm$^{-3}$) for the dust and gas to be well coupled and in thermal equilibrium~\cite{Goldsmith2001}. Thus, we used the gas temperature derived from \tmthc{} as an approximation of the dust temperature. The corresponding temperature derived from high-resolution data for each condensation is presented in Table.~\ref{tab1}.

We use the temperatures derived from ALMA high-resolution and low-resolution \tmthc{} data to calculate the parameters of condensations and their parent disk, respectively. 
The derived condensation gas masses, $M_{\rm gas}$, range from 0.04 to 0.25 \msun{} (Table.~\ref{tab1}), resulting in volume densities of 0.97 $\times$ 10$^{9}$ -- 3.01 $\times$ 10$^{9}$ cm$^{-3}$.  
The column densities are estimated in the range from  4.38$\times 10^{24}$ to 11.47$\times 10^{24}$ cm$^{-2}$. With such high-column-density the condensations  could completely  absorb the central protostellar light.  
We derive a parent disk mass, $M_{\rm disk}$, of 1.87~\msun{}  by subtracting the total condensation gas mass of 0.84~\msun{} from the total parent structure gas mass of 2.71~\msun. 
The estimated surface density distributions are shown in Extended Data Fig.~\ref{fig:Tgas}. 
The derived gas masses and densities should be considered as lower limits because of potential high optical depth and the observations suffered from missing flux in the interferometer data. 
Following refs.~\cite{Li2020,Sanhueza2017}, we adopt uncertainties of 28\%, 23\%, and 10\% for the dust opacity,  gas-to-dust ratio, and continuum flux, respectively. 
The uncertainty in the distance from the trigonometric parallax measurement is around 10\% (refs.\cite{Chibueze2014,Reid2014}). 
We adopt a conservative uncertainty of 30\% for the temperature.  
Consequently, we estimate the uncertainties in the mass, volume density, and surface density are  53\%, 50\%, and 49\%, respectively.

\vspace{0.2em}
\noindent\textbf{Jeans length}

If the fragmentation is governed by Jeans instability, the Jeans length $\lambda_{J}$, which is the separation between fragments, can be calculated as follows~\cite{Kippenhahn2013}
\begin{equation}\label{equ:sblaw_garay}
\lambda_{J} = \sqrt{\frac{\pi \, \sigma^{2}_{\rm eff}}{G\, \rho}} = 0.06\; {\rm pc} \, \left(\frac{\sigma_{\rm eff}}{0.188\; {\rm km\, s^{-1}}} \right) \left(\frac{\rho}{10^5 \; \rm cm^{-3}}\right)^{-1/2}, 
\end{equation}
where $\sigma_{\rm eff}$ is the effective velocity dispersion, $\rho$ is the volume density, and $G$ is the gravitational constant. 
The $\sigma_{\rm eff}$ is equal to the sound speed $c_{\rm s}$ = 0.87 \kms{} for thermal Jeans fragmentation, based on the mean temperature of 216~K. 
The average volume density of the parent structure of the septuple system is $\rho$ = 2.4 $\times\, 10^{8}$~cm$^{-3}$, calculated based on the enclosed mass $M_{\rm cl}$. 
The derived thermal Jeans length is 1168~AU, which is larger than the observed deprojected mean separation of 332~AU for the septuple system.  
Assuming a lower temperature of 50 K and a derived density of $\rho = 2.4 \times 10^{8}$ cm$^{-3}$, the thermal Jeans length is estimated to be $\sim$562~AU. This value should be considered a lower limit, as the density was likely lower and the temperature within the disk could be higher than 50 K  at the early evolutionary stages of the disk according to simulations \cite{Oliva2020}.  The Jeans length derived from the aforementioned assumptions does not fully rule out the possibility of multiple system formation via core fragmentation followed by significant inward migration over a short timescale. However, the overall evidence is more consistent with a disk fragmentation scenario. 
In the turbulent Jeans fragmentation scenario, $\sigma_{\rm eff}$ incorporates both thermal and non-thermal velocity components, $\sigma_{\rm eff} = \sqrt{c_{\rm s}^2 + \sigma_{\rm nth}^2}$, assuming that turbulence acts as an isotropic support.  The measured mean velocity dispersion of \mthc{} is $\sigma_{\rm eff}$ = 2.6 \kms{} for the parent structure, resulting in turbulent Jeans lengths of 3460~AU.   
The maximum recoverable scale of our observations is 3.0\arcsec{}, which is larger than the observed disk diameter of $\sim$1.5\arcsec{}. 
Therefore, the missing flux does not affect our main conclusions.  
If we use the clump averaged volume density of 2 $\times$ 10$^{7}$~cm$^{-3}$  (ref.\cite{Sandell2000}), the derived thermal Jeans length is 4037~AU and the turbulent Jeans length being even larger. 
Although we cannot completely rule out the possibility that some members of the multiple system may have formed externally via core fragmentation and subsequently migrated inward to the circumstellar disk within a short timescale \cite{Guszejnov2023,Lee2019,Kuruwita2023}, the overall observed properties of the septuple system are more consistent with a disk fragmentation scenario.

\vspace{0.2em}
\noindent\textbf{Dynamical mass of the embedded structures}

The \mthc{} $12_{4} -11_{4}$ line, a typical disk tracer, exhibits high signal-to-noise ratios and is relatively isolated from other molecular lines, thus suffering less from confusion caused by line blending.  
We observed a velocity gradient perpendicular to the outflow direction, indicating the presence of a disk kinematic structure toward the parent structure of the septuple system (Fig.~\ref{fig:cont0}c). 
The position-velocity (PV) diagram of \mthc{} further supports that the parent structure is a rotating Keplerian-like disk (Fig.~\ref{fig:cont0}d). 
This is consistent with simulations showing that, although the fragments are presented within the disk (Fig.~\ref{fig:simul}), the gas kinematics of the disk retain a Keplerian-like profile \cite{Ahmadi2019}. 
The fragments can affect the local gas kinematics, as evidenced by variations in the high-resolution velocity map (Fig.~\ref{fig:vlsr}), but they do not significantly alter the overall   kinematic profile of the parent disk at the current stage. 

We employ the {\tt 3DBarolo} code to fit the disk kinematics by modeling the three-dimensional (3D) position-position-velocity image cube \cite{DiTeodoro2015}, which uses a series of concentric tilted rings to derive the rotation velocities, inclination angle, and position angle of the disk by assuming a circular geometry. 
We applied {\tt 3DBarolo} to the low-resolution \mthc{} $12_{4} -11_{4}$ image cube for characterizing the properties of the disk, as the SNR of the high-resolution \mthc{} image is insufficient for robustly determining the disk parameters. 
Supplementary Fig.~\ref{fig:model} shows kinematic maps of the observations, model, and corresponding residuals for intensity, velocity, and velocity dispersion.  
The best-fits results in an inclination angle of 26.3$^{\circ}$ (0$^{\circ}$ represents a face-on view) and a position angular of 292.5$^{\circ}$ (from north to east), with a few degrees of variations throughout different rings. 
The best-fit rotation velocities are shown as green circles in Fig.~\ref{fig:cont0}d. 
A power-law fit to the rotation velocities results in a relation of 
$V_{\rm rot} \propto r^{-0.53 \pm 0.09}$ (green curve in Fig.~\ref{fig:cont0}d), which is consistent with Keplerian rotation. 
The enclosed mass of the disk can be estimated under the assumption of Keplerian rotation, $V_{\rm rot} \propto r^{-0.5}$. 
The derived enclosed mass is $M_{\rm cl}$ = 7.6 $\pm$ 1.4~\msun{}, which is close to 10~\msun{} estimated from the relatively lower resolution and lower sensitivity observations reported by ref.~\cite{Hunter2014}. 
 
We present the high-resolution observation PV diagram of \mthc{} in Supplementary Fig.~\ref{fig:pvhigh}, which shows the Keplerian-like kinematic and its radial velocity profile broadly consistent with the mass of 7.6~\msun{} derived from the low-resolution data. 
Unfortunately, the SNR of high-resolution \mthc{} image does not permit reliable constraints on the disk parameters.

\vspace{0.2em}
\noindent\textbf{Calculation of the Toomre $Q$ parameter}

The Toomre  $Q$ parameter can be used to study the gravitational stability of the disk~\cite{Toomre1964}: 
\begin{equation}\label{equ:TQ}
Q = \frac{c_{\rm s} \kappa}{\pi G \sum}, 
\end{equation}
where $c_{\rm s}$ is the sound speed, $G$ is the gravitational constant,  
$\sum$ is the disk surface density that is computed from disk mass of 2.71~\msun{}  with Equation~\ref{surface}, and $\kappa$ is epicyclic frequency that is equivalent to the angular velocity $\Omega$ = $\sqrt{GM/r^3}$ in the case of a  Keplerian disk. 
The disk becomes unstable against axisymmetric gravitational stability if $Q <$ 1--2  (refs.~\cite{Kratter2016,Durisen2007}). 
The $\kappa$ is derived from enclosed mass of $M_{\rm cl}$  = 7.6~\msun{}, which is  estimated by the Keplerian model, 
and temperature derived from low-resolution and high-resolution \tmthc{} data to calculate the $Q$ maps for the disk.  
The results are presented in Fig.~\ref{fig:TnQ}a and Fig.~\ref{fig:TnQ}b for low-resolution and high-resolution maps, respectively.  
The derived $Q$ maps consistently show a decrease in $Q$ with increasing disk radius (Extended Data Fig.~\ref{fig:radTnQ}). 
As expected for disk fragmentation, we find low $Q$ values around condensations C2, C3, C4, C5, C6, and C7 (Table.~\ref{tab1}),  where regions have likely already collapsed to form protostars, while higher $Q$ values appear toward the central protostar (C1). The protostellar radiation  can prevent further fragmentation by raising the temperature and thus stabilizing the central region of the disk.  
High $Q$ values are revealed to the northeast and southwest of C1, coinciding spatially  with the direction of molecular outflows.    
The main sources of uncertainty in the derived Toomre $Q$ parameters originate  from the sound speed, surface density, and epicyclic frequency. 
The epicyclic frequency uncertainty is 18\% resulting from $M_{\rm cl}$ and distance. 
The sound speed uncertainty, propagated from the temperature, is 15\%.    
Then the uncertainty in the Toomre $Q$ parameter is estimated to be 54\%.

\vspace{0.2em}
\noindent\textbf{Virial parameter of condensations}

The virial parameter, $\alpha_{\rm vir} = M_{\rm vir}/M_{\rm tot}$, is frequently used to evaluate the gravitational state of dense objects \cite{Bertoldi1992}. Nonmagnetized condensations with $\alpha_{\rm vir} <$ 1, $\alpha_{\rm vir} \sim$ 1, and $\alpha_{\rm vir} <$ 2  are considered to be in  gravitationally unstable, hydrostatic equilibrium, and gravitationally bound \cite{Bertoldi1992}, respectively.  
The virial mass, $M_{\rm vir}$, can be calculated following \cite{Bertoldi1992}:  
\begin{equation}
\label{equ:Mvir}
M_{\rm vir}  = \frac{5}{a \beta} \frac{\sigma^{2} R}{G},
\end{equation} 
where $R$ is the condensations radius, $a=(1-b/3)/(1-2b/5)$ is the correction factor for a power-law density profile $\rho \propto R^{-b}$, $\beta = ({\rm arcsin}\; e)/e$ is the geometry factor \cite{Fall1983,Li2013}, $\sigma$ is the velocity dispersion of the gas in the condensation. 
The eccentricity $e$ is calculated by the axis ratio of the dense structure, $e \, = \, \sqrt{1 - f_{\rm int}^{2}}$. 
Considering  projection effects and given that the dense cores are likely prolate ellipsoids,  the intrinsic axis ratio, $f_{\rm int}$, is estimated from observed axis ratio, $f_{\rm obs}$,  with $f_{\rm int}  = \frac{2}{\pi} \,f_{\rm obs} \, \mathcal{F}_{1}(0.5, 0.5, -0.5, 1.5, 1, 1-f_{\rm obs}^{2})$ (ref.~\citep{Fall1983}), where $\mathcal{F}_{1}$ is the Appell hypergeometric function of the first kind. We adopted a typical density profile index of b = 2 for all condensations. 
The velocity dispersion is calculated by $\sigma$ = $\sigma_{\rm tot}$ - $\sigma_{\rm rot}$, where $\sigma_{\rm tot}$ is the total velocity dispersion, and $\sigma_{\rm rot}$ is the disk rotation induced velocity dispersion. The total velocity dispersion is given by $\sigma_{\rm tot}$ = $\sqrt{\sigma_{\rm obs}^2 - (\triangle{\rm ch}/2\sqrt{\rm 2 ln2})^2 - \sigma_{\rm th}^2 + c_{\rm s}^2}$, where $\sigma_{\rm obs}$ is measured from a Gaussian fit to the condensation-averaged \mthc{} $12_{4} -11_{4}$ spectra using the high-resolution data, $\sigma_{\rm th}$ is the \mthc{}  thermal velocity dispersion, and $\triangle_{\rm ch}$ is the channel width. 
The high-resolution Keplerian model is not available due to the low SNR of the high-resolution  \mthc{} image. Therefore, we adopt the values from the low-resolution Keplerian model at the corresponding condensation positions as an approximation for $\sigma_{\rm rot}$ (Supplementary Fig.~\ref{fig:model}).

In the simulations, the fragment gas masses are estimated to range from 0.03 to 0.23 \msun{},  using the same procedures as in the observations and assuming a dust temperature of 200~K.  
The total mass of each fragment is approximately 1 \msun{}, calculated by direct integration over the density according to ref.~\cite{Oliva2020}. 
In addition, the average dust optical depth of the fragments is around 10 in most of cases, assuming a $\kappa_{\rm 1.3 mm}$ of 0.9 cm$^{2}$ g$^{-1}$. 
These suggest that the calculated gas mass is likely an order of magnitude lower than the actual total mass. Therefore, we assume the total mass of the condensations is 10 times of their gas mass, i.e., $M_{\rm tot}$ = 10 $M_{\rm gas}$, which is a reasonable assumption given that the observations share similar properties with the simulations. 
The derived virial masses are between 0.08 and 2.5 \msun{}. 
The virial parameter, $\alpha_{\rm vir} = M_{\rm vir}/M_{\rm tot}$, ranges from 0.1 to 1.8 (Table.~\ref{tab1}), indicating that the condensations are gravitationally bound.

\vspace{0.2em}
\noindent\textbf{Boundedness of the septuple system}

We compute the gravitational potential energy ($W_{\rm i}$) and kinetic energy  ($E_{i}$) of each condensation to asses whether it is gravitationally bound to the system. 
The $W_{\rm i}$ and $E_{i}$ values are calculated by 
\begin{equation}
\label{equ:grav}
W_{\rm i}  = - \frac{G \, M_{\rm cl} \, m_{\rm i}}{ R_{\rm i0} },
\end{equation} 

\begin{equation}
\label{equ:kine}
E_{i}  = \frac{1}{2} m_{i} (\rm V_{i} - V_{\rm disk})^2,
\end{equation} 
where $M_{\rm cl}$ is the enclosed mass,  $G$ is the gravitational constant, 
$m_{i}$ is the gas mass of condensation $i$,  $R_{i0}$ is the distance from the condensation to the center of the disk, V$_{i}$ is the (line-of-sight) velocity of condensation $i$, and V$_{\rm disk}$ = -3~\kms{} is the velocity of the disk.

A condensation with $E_{i}/|W_{\rm i}| \, <$ 1 is considered to be bound to the system. 
To mitigate the optical depth effect and confusion from the gas in outer layers along the line of sight, we measure the centroid velocity of each condensation and parent disk 
by performing a Gaussian fit to the condensation-averaged and disk-averaged \tmthc{} $13_{3}-12_{3}$ spectra, respectively.  
We use the condensation gas mass $M_{\rm gas}$ to calculate the $E_{i}$ for each condensation, expect for C1, which is located at the center of the disk. 
Following ref.~\cite{Li2024,Pineda2015}, we estimated the $E_{i}/|W_{M_{\rm cl}}|$ with four different methods (Extended Data Fig.~\ref{fig:EiWi}), including (1) line-of-sight velocity difference and on-sky separation  (refer to one-dimensional, 1D; black symbols),
(2) three-dimensional (3D) velocity difference ($\sqrt{3}$ times the line-of-sight velocity difference) and on-sky separation (red symbols),
(3) line-of-sight velocity difference and deprojected separation the inclination angle of 26.3$^{\circ}$ (blue symbols),
and (4) 3D velocity difference and deprojected separation (orange symbols).

As shown in Extended Data Fig.~\ref{fig:EiWi},  the $E_{i}/|W_{M_{\rm cl}}|$ is lower than 1 for all condensations (Table.~\ref{tab1}). 
Even if we include the central protostar mass of each condensation and assume this to be 10 times the condensation gas mass, the resulting $E_{i}/|W_{M_{\rm cl}}|$ is $<$0.6.  
The results indicate that the septuple system is gravitationally bound at the present stage.

\vspace{0.2em}
\noindent\textbf{Gas properties around the septuple system}

As shown in Fig.~\ref{fig:vlsr} and Extended Data Fig.~\ref{fig:mom02}, the   presence of a velocity gradient, broadened linewidths, and enhanced intensity across the condensations indicates possible small-scale accretion disks  surrounding them. 
Such small-scale secondary disks embedded in the primary disk have been found in simulations of multiplicity formation formed through disk fragmentation \cite{Oliva2020}.
However, we cannot completely rule out the possibility that the variation in line emission may be attributed to small-scale gas fluctuations and/or gas flow processes.  
Unfortunately, search for the secondary disks in these condensations is limited by the sensitivity, angular and spectral resolutions of our data.  
Higher resolution and sensitivity observations are needed to confirm the presence of individual secondary disks around the condensations.

\vspace{0.2em}
\noindent\textbf{Simulations of disk fragmentation and multiplicity formation}

The simulation models the collapse of a molecular cloud with an initial mass of 200  $M_{\odot}$ contained within a radius of 0.1 pc, and focuses on the formation of a central massive protostar and the evolution and fragmentation of its surrounding accretion disk. The density is initially distributed with a profile $\rho(r) \propto r^{-1.5}$ and the initial rotation rate is determined by a rotational-to-gravitational energy ratio of 5\% and distributed according to a profile $\Omega(r) \propto r^{-0.75}$. 
These initial values were motivated by typical properties of dense cores in massive star forming regions \cite{Beuther2018}, but a qualitatively similar picture is obtained by considering variations in the initial angular momentum distribution of the cloud \cite{Meyer2017}. 
The simulation achieves extremely high spatial resolution using a spherical coordinate grid, with $268 \times 81 \times 256$ cells in the radial, polar, and azimuthal directions, respectively. This configuration allows resolving the disk down to 0.75 au at the inner rim (30 au) and 25 au at the outer edge ($\sim$1000~au), providing over a million grid cells to represent the disk. More details on the simulation can be found in ref.\cite{Oliva2020}.
 
The hydrodynamic evolution of the gas and dust is modeled using the Pluto code \cite{Mignone2007,Mignone2012}, including self-gravity through the module Haumea and radiation transport using the module Makemake \cite{Kuiper2011,Kuiper2010}. The radiation transport treatment takes into account stellar irradiation and absorption and (re-)emission by the gas and dust, and uses the protostellar evolution tracks produced by ref.\cite{Hosokawa2009}.
 
After its formation, the disk undergoes gravitational instabilities, leading to the formation of fragments and spiral arms (Fig.~\ref{fig:simul}). The fragments interact, merge and develop disks with spiral arms on their own. To create synthetic observations of this evolving structure, we perform a radiative transfer (RT) post-processing simulation, using the disk configuration approximately 10 kyr after the onset of collapse. At this point, the central star has grown to $\sim$6.68 $M_{\odot}$, though it is not included in the RT simulation. 
This time instant was chosen for the comparison because of the resemblance of the spatial distribution of the fragments, the mass of the central massive protostar and the size of the disk. 
The densities and/or temperatures of some fragments have reached the criteria of the first Larson core according to the simulations \cite{Oliva2020}.   
In the simulation, the disk contains between 4 and 9 fragments (including the central protostar) at times between 8 and 12 kyr of evolution.

For the synthetic observations, we use the open-source software RADMC-3D  to generate dust continuum emission at 1.3 mm and molecular line emissions for the \mthc{} $12_{4} -11_{4}$ transition \cite{Dullemond2012}. Key parameters for the RT simulation, such as gas and dust densities, temperatures, and velocities, are taken directly from the hydrodynamic simulation snapshot. Dust temperatures are assumed to be in thermal equilibrium with the gas, and dust opacities follow the work of ref.\cite{Laor1993}. The $\rm{CH_3CN}$ abundance is determined based on temperature thresholds, following guidelines from refs.\cite{Collings2004,Gerner2014}.
 
We model a 1000 AU $\times$ 1000 AU region around the protostar with a spatial resolution of 1 AU per pixel. The disk is assumed to be inclined at 27$^{\circ}$  (0$^{\circ}$ for face-on) and is located at a distance of 1.3 kpc  (Fig.~\ref{fig:simul}). This setup allows the post-processing to generate synthetic data cubes that mimic ideal telescope observations.
We also show an example of simulation continuum image that contains 7 fragments within the disk in Extended Data Fig.~\ref{fig:simu7} for reference.

\vspace{0.2em}
\noindent\textbf{Summerizng the measured masses from the observations}

The estimated condensation gas masses, $M_{\rm gas}$, range from 0.04 to 0.25 \msun, derived from the high-resolution 1.3~mm dust continuum image. 
The disk mass, $M_{\rm disk}$ = 1.87 \msun, is estimated from the low-resolution 1.3~mm dust continuum image. 
The enclosed mass, $M_{\rm cl}$ = 7.6 \msun\ is determined by fitting a Keplerian model to the disk rotation velocities.  
The protostellar mass is estimated as $M_{\ast}$ =  $M_{\rm cl}-M_{\rm disk}-\sum M_{\rm gas}$ = 4.89~\msun.

\vspace{1em}
\noindent\textbf{Data availability}

This paper makes use of the following ALMA data: ADS/JAO.ALMA \#2021.1.00713.S and \#2022.1.00671.S.  ALMA is a partnership of ESO (representing its member states), NSF (USA) and NINS (Japan), together with NRC (Canada), NSTC and ASIAA (Taiwan), and KASI (Republic of Korea), in cooperation with the Republic of Chile. The Joint ALMA Observatory is operated by ESO, AUI/NRAO and NAOJ.
The raw data are available at \url{https://almascience.nao.ac.jp/aq}  by setting the observation code. 
Owing to their size, the reduced data used for this study are available from the corresponding authors upon reasonable request. 

\vspace{0.2em}
\noindent\textbf{Code availability}

The ALMA data were reduced using CASA versions 6.4.1-12 that are available at \url{https://casa.nrao.edu/casa_obtaining.shtml}. 
The {\tt pvextractor} package was used to make position velocity diagram, which  is available at  \url{https://pvextractor.readthedocs.io}. 
The disk kinematics fitting package of {\tt 3DBarolo} is available at \url{https://editeodoro.github.io/Bbarolo/}. 
The XCLASS is available at \url{https://xclass.astro.uni-koeln.de}.

\clearpage

\setcounter{figure}{0}
\renewcommand{\figurename}{Extended Data Fig.}
\renewcommand{\figureautorefname}{Extended Data Fig.}

\renewcommand{\tablename}{Extended Data Table.}


\begin{figure}[!h]
\centering
\includegraphics[width=1\textwidth]{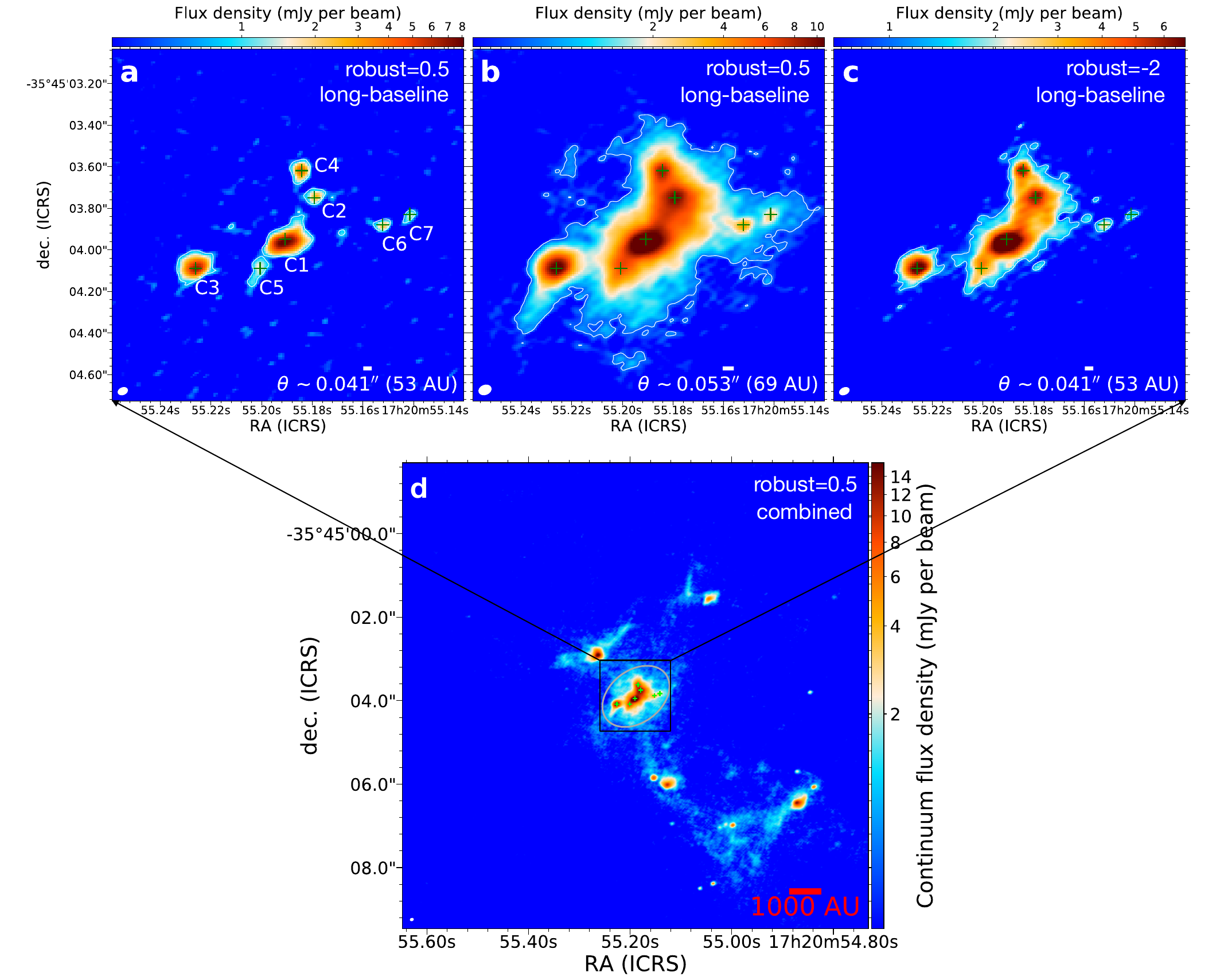}
\caption{
\textbf{Continuum images at different angular resolutions.} 
\textbf{(a)}--\textbf{(c)} show 1.3~mm continuum images with an angular resolutions of roughly 0.041\arcsec, 0.053\arcsec\ and 0.041\arcsec, respectively.  
The continuum image is generated using the uv range of $>$1.3 km in the long-baseline data for \textbf{(a)}, while the full long-baseline dataset is used for \textbf{(b)} and \textbf{(c)}. 
The septuple protostar system is shown as green pluses. 
The white contour indicates the 5$\sigma$ level for each image, where $\sigma$ are 0.15, 0.11, and 0.23~mJy  per beam for \textbf{(a)}, \textbf{(b)}, and \textbf{(c)}, respectively. 
\textbf{(d)} zoom-out view of the septuple system in the 1.3~mm continuum image. The image is produced using the combined dataset, achieving an angular resolutions of roughly 0.06\arcsec. 
The white ellipses in the bottom-left corner of each panel denote the synthesized beam of continuum images.    
The linear scale bar is displayed in the bottom-right corner of panels.
}
\label{fig:cont1}
\end{figure}
\begin{figure}[!h]
\centering
\includegraphics[width=1\textwidth]{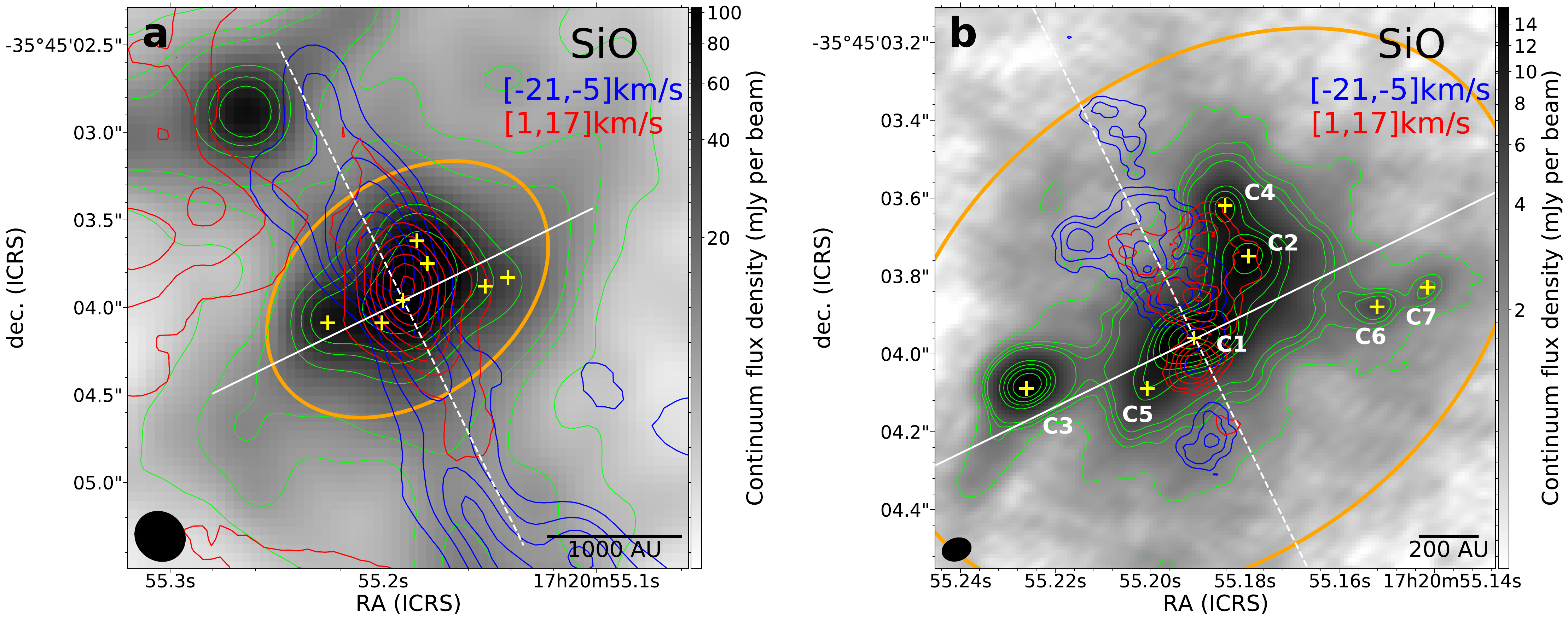}
\caption{
\textbf{Outflows emission.} 
\textbf{(a)} and \textbf{(b)} SiO redshifted (red line) and blueshifted (blue line) emission overlaid on 1.3~mm continuum (green line) for  ALMA low-resolution and high-resolution data, respectively.  
Contour levels start  at  3$\sigma_{\rm rms}$ and increase in step of 2$\sigma_{\rm rms}$ for the low-resolution image and 1$\sigma_{\rm rms}$  for the high-resolution image. 
The $\sigma_{\rm rms}$ are 0.05 and 0.017 Jy per beam km s$^{-1}$ for low-resolution and high-resolution SiO data, respectively.  
The yellow ellipse shows the parent structure of the septuple system.  
The lime contours show the continuum image, with contour levels are [5, 15, 25, 35, 80, 120, 160, 260]$\times \sigma$ with $\sigma$ of 0.36~mJy per beam for the low-resolution image, and [15, 25, 30, 35, 70, 85, 100, 120, 160, 200] $\times \sigma$ with $\sigma$ of 0.12~mJy per beam for the high-resolution image. 
The black ellipses in the bottom-left corner of each panel denote the synthesized beam of corresponding images.  
The linear scale bar is displayed in the bottom-right corner of the panels.
}
\label{fig:kine}
\end{figure}
\begin{figure}[!h]
\centering
\includegraphics[width=1\textwidth]{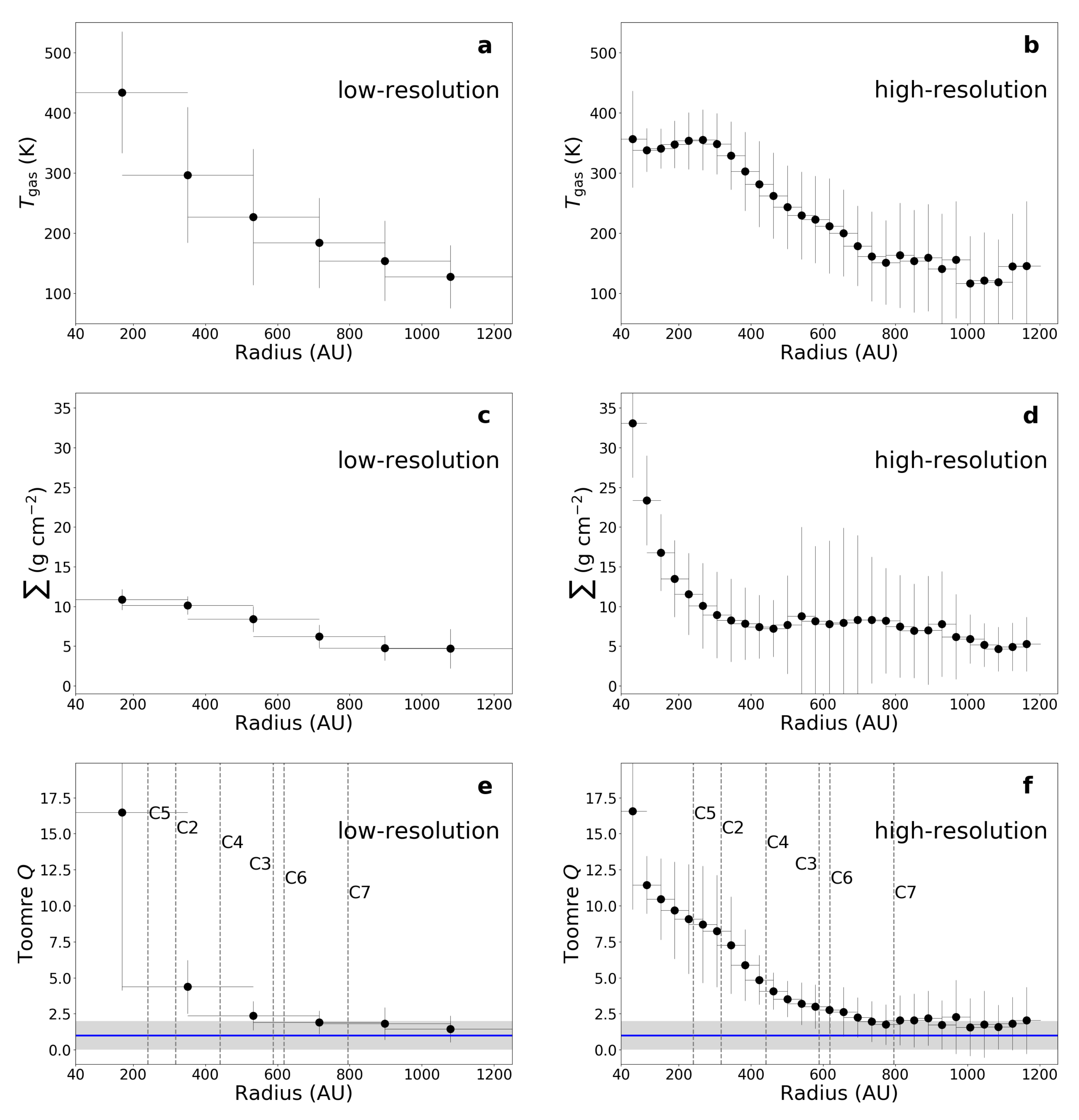}
\caption{
\textbf{Gas temperature ($T_{\rm gas}$), surface density ($\sum$), and Toomre's $Q$ radius profiles.} 
The plots show $T_{\rm gas}$, $\sum$ and Toomre's $Q$ averaged value in a series of circular annuli centered on C1 with the annulus width of half the beam size. 
The error bars correspond to the standard deviation inside each annulus. 
The left and right columns show the results derived from low-resolution  and high-resolution maps, respectively. 
The grey shaded area shows the $Q$ values between 1 and 2, where the disk is expected to be gravitationally unstable. The blue solid line marks $Q$ = 1. 
The vertical dashed lines indicate the separation of condensations from  C1. 
}
\label{fig:radTnQ}
\end{figure}
\begin{figure}[!h]
\centering
\includegraphics[width=1\textwidth]{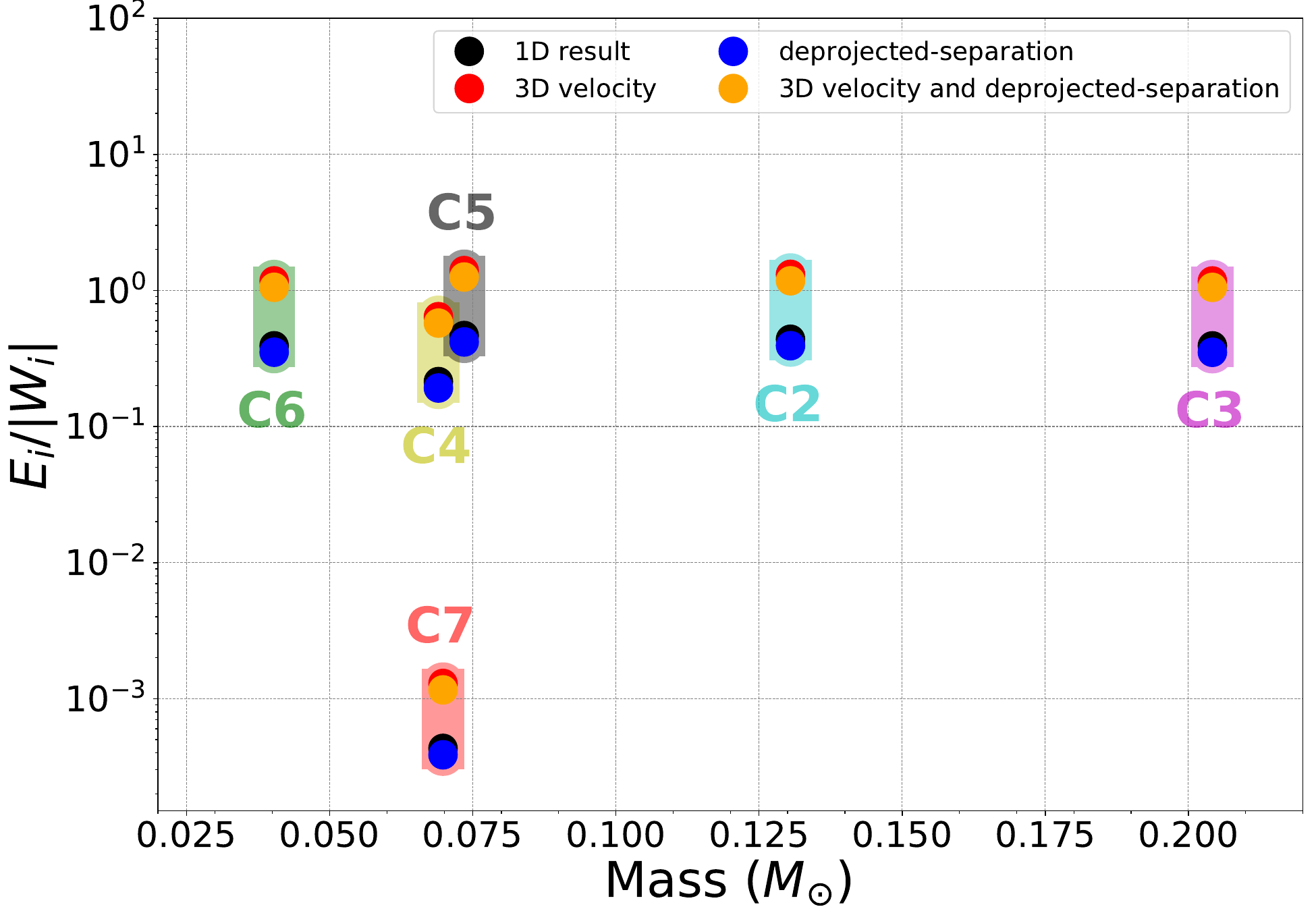}
\caption{
\textbf{The kinetic-to-gravitational energy ratio for the septuple system.} 
The condensations are indicated with different color shadows. 
The multiple systems are considered to be gravitationally bound if  the kinetic-to-gravitational energy ratio is below unity. 
The ratio of $E_{i}/|W_{\rm i}|$ is $\lesssim$  1 for all condensations, indicating that the septuple system is gravitationally bound. 
}
\label{fig:EiWi}
\end{figure}
\begin{figure}[!h]
\centering
\includegraphics[width=1\textwidth]{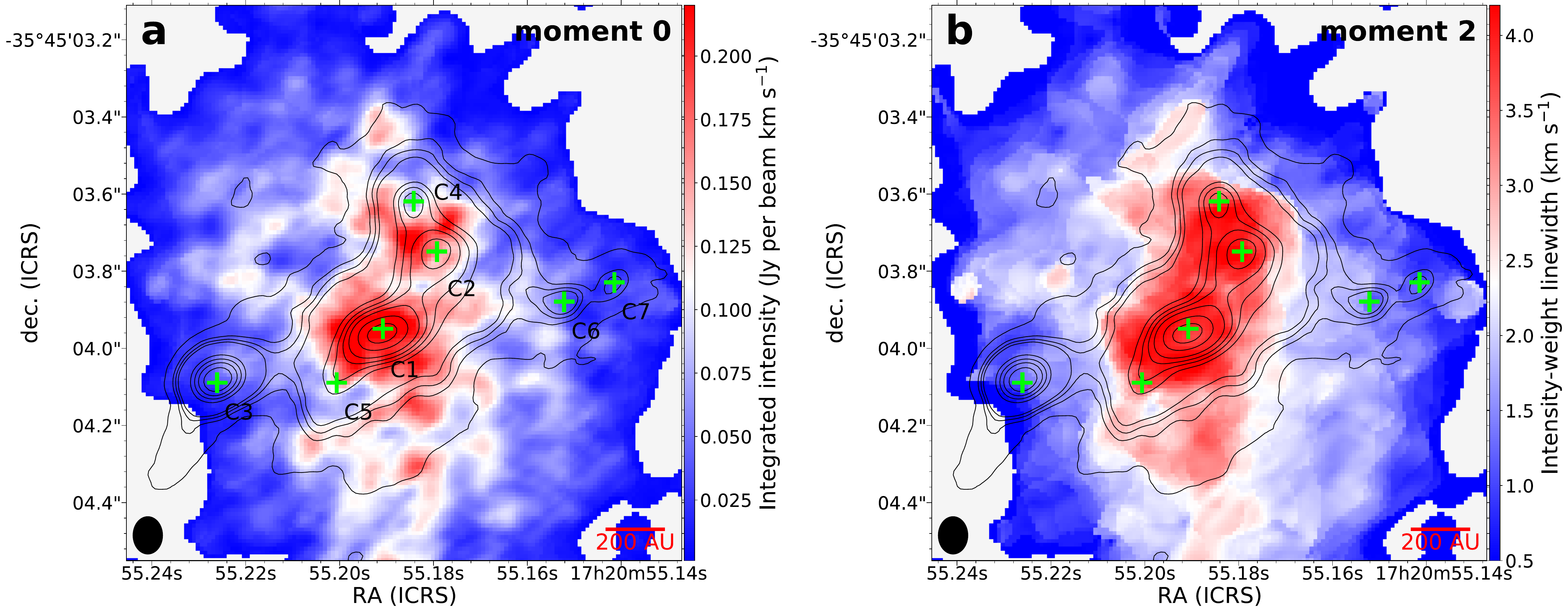}
\caption{
\textbf{Intensity and linewidth distributions of CH$_{3}$CN within the disk.} 
\textbf{(a)} and \textbf{(b)} The velocity-integrated intensity (moment 0) and intensity-weighted linewidth (moment 2)  maps of CH$_{3}$CN $12_{4} -11_{4}$ derived from high-resolution data. 
The black contours show the ALMA high-resolution 1.3~mm continuum image, with contour levels at  [15, 25, 30, 35, 70, 85, 100, 120, 160, 200] $\times \sigma$ with $\sigma$ of 0.12~mJy per beam. 
The synthesized beam of \mthc{} image is presented in the bottom-left corner as a black ellipse.  The linear scale bar is displayed in the bottom-right corner of the panels.
}
\label{fig:mom02}
\end{figure}
\begin{figure}[!h]
\centering
\includegraphics[width=1\textwidth]{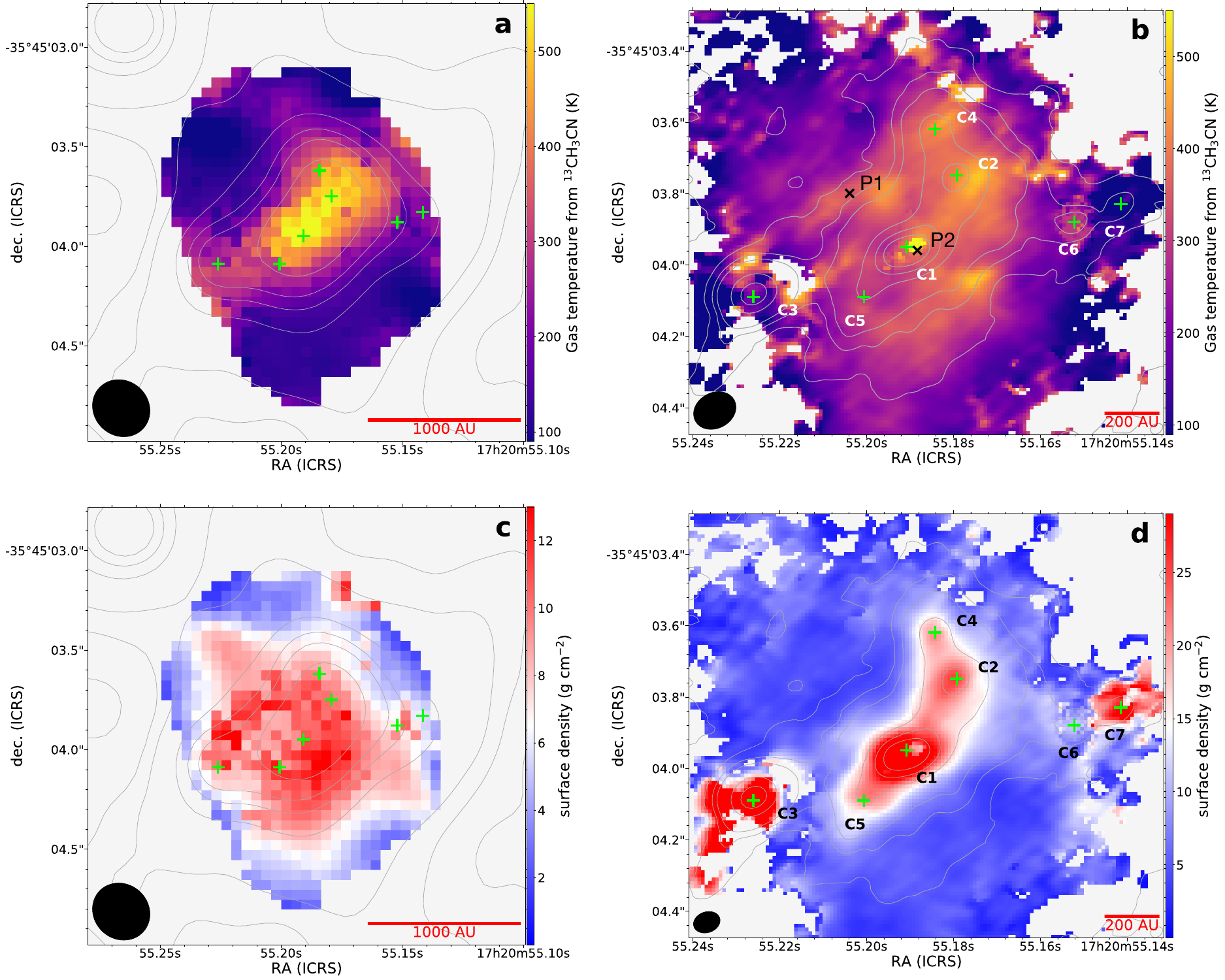}
\caption{
\textbf{Gas temperature and surface density maps.} 
\textbf{(a)} and \textbf{(b)} The gas temperature maps derived from the low-resolution and high-resolution \tmthc{} data, respectively. 
The green pluses indicate the condensations. 
The black crosses mark the positions used to extract the example spectra of \tmthc{}. 
\textbf{(c)} and \textbf{(d)} The surface density distributions derived from low-resolution and high-resolution data, respectively.  
The grey contours show the ALMA high-resolution 1.3~mm continuum image, with contour levels at [5, 15, 25, 35, 80, 120, 160, 260]$\times \sigma$ with $\sigma$ of 0.36~mJy per beam for the low-resolution image, and [15, 25, 30, 35, 70, 85, 100, 120, 160, 200] $\times \sigma$ with $\sigma$ of 0.12~mJy per beam for the high-resolution image. 
The black ellipses in the bottom-left corner of each panel denote the synthesized beam of corresponding images.  The linear scale bar is presented in the bottom-right corner of the panels.  
}
\label{fig:Tgas}
\end{figure}

\begin{figure}[!h]
\centering
\includegraphics[width=0.7\textwidth]{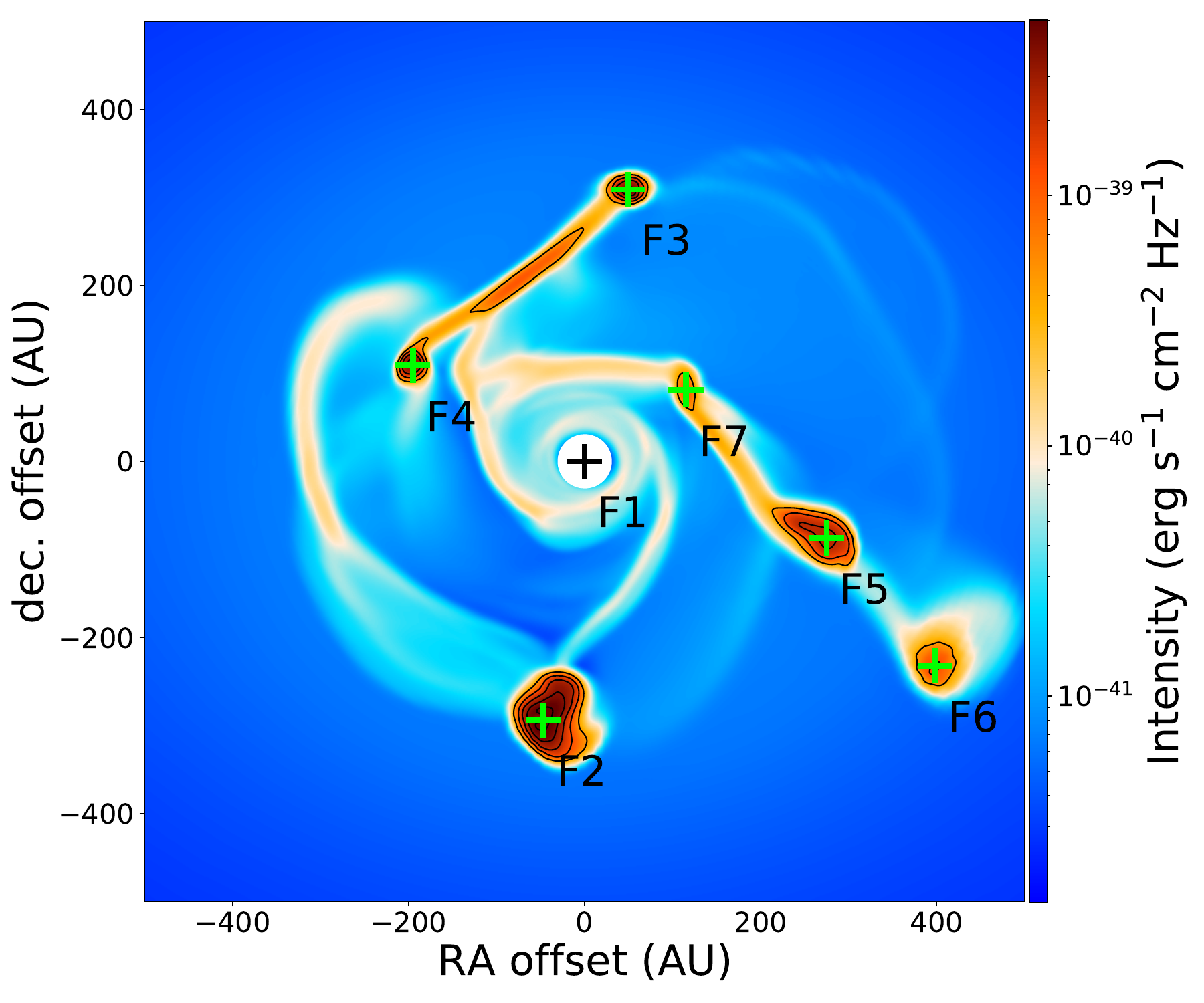}
\caption{
\textbf{1.3~mm continuum image of simulations.} 
The 1.3~mm continuum image of simulations is inclined by 27$^{\circ}$ at a  distance of 1.3~kpc. The simulations image reveals a septuple system formed via disk fragmentation. The central source is masked out. The green pluses indicate the fragments within the disk. 
The black contour levels are [0.45, 1.18, 2.25, 3.62, 5.35, 6.59, 7.83, 9.08]$\times 10^{-39}$ erg s$^{-1}$ cm$^{-2}$ Hz$^{-1}$. 
}
\label{fig:simu7}
\end{figure}

\clearpage

\setcounter{figure}{0}
\renewcommand{\figurename}{Supplementary Fig.}
\renewcommand{\figureautorefname}{Supplementary Fig.}



\begin{figure}[!h]
\centering
\includegraphics[width=1\textwidth]{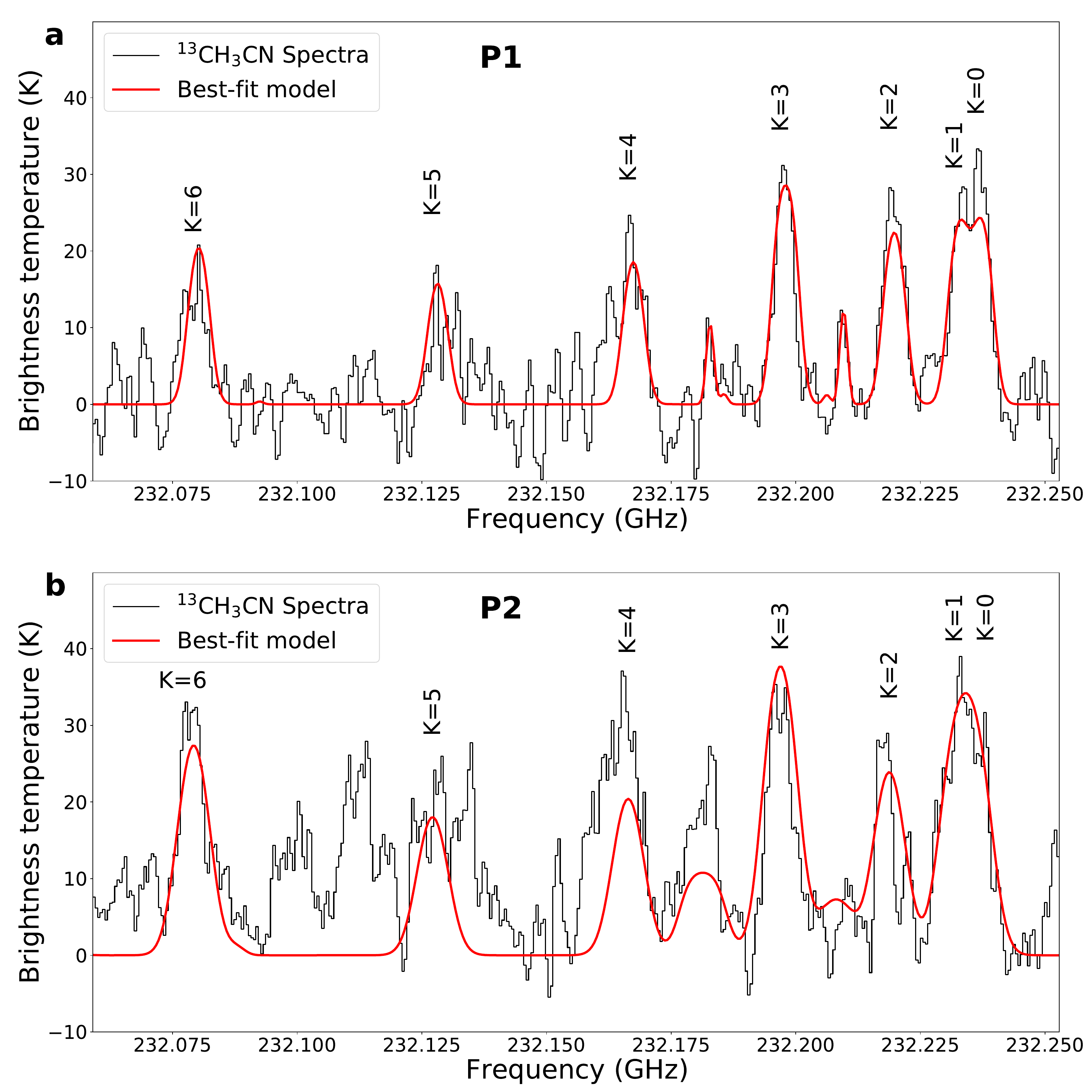}
\caption{
\textbf{Example \tmthc{} spectra.} 
The observed spectra of \tmthc{} extracted from  two positions using the high resolution data.  The positions P1 and P2 are marked with black crosses in high-resolution gas temperature map (Extended Data Fig.~6). 
The result of {\tt XCLASS} best fit is overlaid on the spectrum as red line.
The CH$_{3}$COOH, CH$_{3}$OCHO, and CH$_{3}$OCH$_{3}$ emission are also included to improve the fit. 
}
\label{fig:13CH3CN}
\end{figure}
\begin{figure}[!h]
\centering
\includegraphics[width=0.88\textwidth]{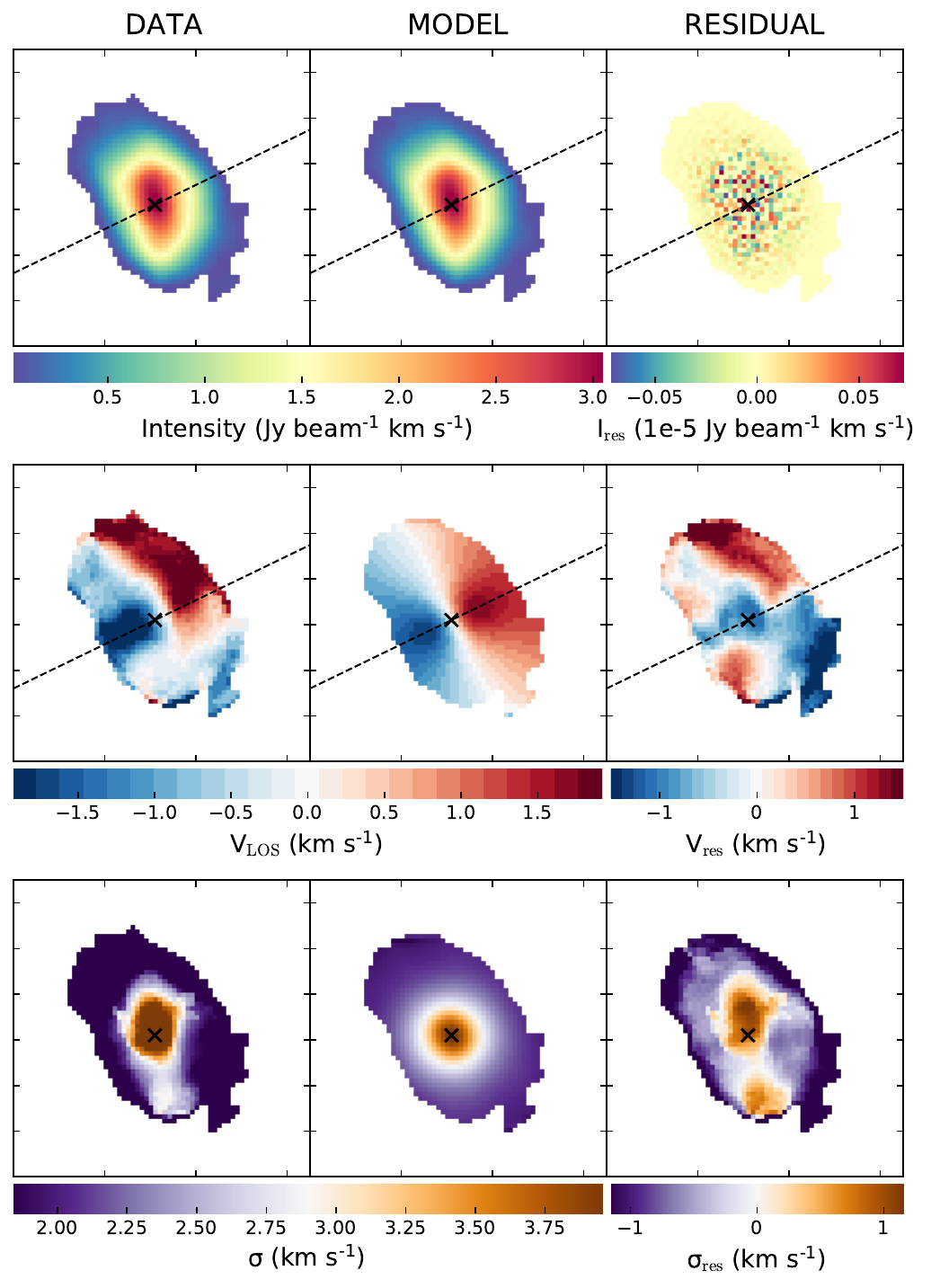}
\caption{
\textbf{Kinematic maps of the observations and model.} 
Left, middle, and right columns show the observations, model, and  corresponding residuals, respectively, for intensity (top row), velocity (middle row), and velocity dispersion (bottom row). The black cross marks the position of C1,  which is used as the disk center for model fitting. 
}
\label{fig:model}
\end{figure}
\begin{figure}[!h]
\centering
\includegraphics[width=0.7\textwidth]{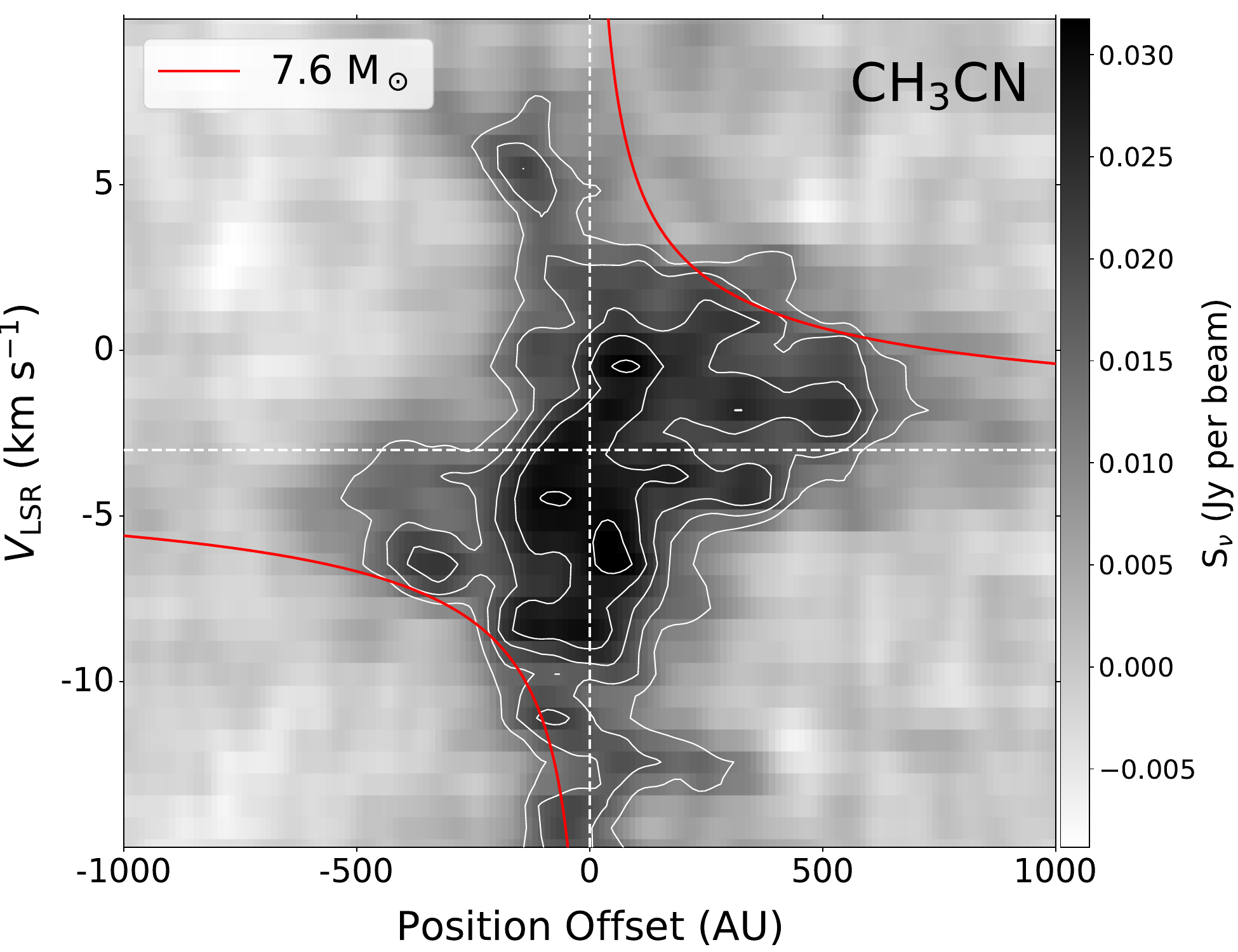}
\caption{
\textbf{PV diagram of high-resolution CH$_{3}$CN emission.} 
PV diagram for high-resolution  CH$_{3}$CN $12_{4} -11_{4}$ emission extracted along the white solid line as indicated in Fig.~3a. 
The red solid line shows the Keplerian velocity curve corresponding to a mass of 7.6 \msun{}. 
Contour levels  start at  5$\sigma_{\rm rms}$ and increase in step of 2$\sigma_{\rm rms}$, where $\sigma_{\rm rms}$ is 2.4 mJy per beam.   
}
\label{fig:pvhigh}
\end{figure}


\clearpage


\end{document}